\documentclass[aps,pre,reprint,groupedaddress,showpacs]{revtex4-1}
\usepackage{graphicx}
\usepackage{dcolumn}
\usepackage{bm}
\usepackage{multirow}
\usepackage{epstopdf}
\usepackage{amsfonts}
\usepackage{amsmath}
\usepackage{setspace}
\usepackage[usenames]{color}
\usepackage{soul}

\begin{document}
\title{\textbf{\Large Traveling chimera patterns in two-dimensional neuronal network}}
\author{Ga\"el R. Simo$^1$}
\author{Patrick Louodop$^1$}
\author{Dibakar Ghosh$^2$}\email{diba.ghosh@gmail.com}
\author{Thierry Njougou$^1$}
\author{Robert Tchitnga,$^{1,3}$}
\author{Hilda A. Cerdeira$^{4,5}$}
\affiliation{$^1$Research Unit Condensed Matter, Electronics and Signal Processing, Department of Physics, Faculty of Science, University of Dschang, P.O. Box 67 Dschang, Cameroon\\
	$^2$Physics and Applied Mathematics Unit, Indian Statistical Institute, 203 B. T. Road, Kolkata-700108, India\\
	$^3$Institute of Surface Chemistry and Catalysis, University of Ulm,Albert-Einstein-Allee 47, 89081 Ulm, Germany\\
	$^4$S\~ao Paulo State University (UNESP), Instituto de F\'{i}sica Te\'{o}rica, Rua Dr.Bento Teobaldo Ferraz 271, Bloco II, Barra Funda, 01140-070 S\~ao Paulo, Brazil\\
	$^5$Epistemic, Gomez $\&$ Gomez Ltda. ME, Av. Professor Lineu Prestes 2242,Cietec, 
			Sala 244, 05508-000 S\~ao Paulo, Brazil}

\date{\today}

\begin{abstract}

We study the emergence of the traveling chimera state in a two-dimensional network of Hindmarsh-Rose burst neurons with the mutual presence of local and non-local couplings. We show that in the unique presence of the non-local chemical coupling modeled by a nonlinear function, the traveling chimera phenomenon occurs with a displacement in both directions of the plane of the grid. The introduction of local electrical coupling shows that the mutual influence of the two types of coupling can, for certain values, generate traveling chimera, imperfect-traveling, traveling multi-clusters, and alternating traveling chimera, ie the presence in the network under study, of patterns of coherent elements interspersed by other incoherent elements in movement and alternately changing their position over time. The confirmation of the states of coherence is done by introducing the parameter of instantaneous local order parameter in two dimensions. We extend our analysis through mathematical tools such as the Hamilton energy function to determine the direction of propagation of patterns in two dimensions.\\

\end{abstract}


\maketitle

\section{Introduction}

The brain is a complex system made up of millions of neurons that interact with each other through synaptic connections. There are two types of synapses:  chemical and electrical. Communication at the level of chemical synapses occurs through the release of neurotransmitters following an action potential, thus inducing an electrical modification of the postsynaptic neuron. The electrical synapse, for its part, allows a continuous and reciprocal interaction by the direct passage of ions through channels that join the membranes of two adjacent neurons. This explains why the synaptic current is proportional to the difference in membrane potentials that it tends to equalize. These electrical synapses constitute a minority mode of communication compared to chemical synapses. However, their presence and importance has been revealed in some neuronal systems such as the cortex, hippocampus, lower olive, striatum and retina of mammals \cite{connors2004electrical, bennett2004electrical}.

\par Neuronal synchronization is characterized by the presence of temporal correlations between the activities of neurons. Synchronized neuronal activities at various spatial and temporal scales are unmistakably manifested in the central neuronal system. This neuronal synchronization is often associated with the emergence of rhythmic brain activity, the frequencies of which span a wide spectrum, from a few units to hundreds of hertz. Recently, thanks to genetic manipulations in the cerebral cortex \cite{deans2001synchronous, blatow2003novel}, in the hippocampus \cite{hormuzdi2001impaired, buhl2003selective} and in the lower olive \cite{long2004small}, the role of electrical synapses in the emergence of rhythmic activities has been demonstrated. In fact the idea of the passive properties of electrical synapses which would tend to bring the potentials of two coupled neurons closer together and therefore synchronize them, is based on the hypothesis that the coupling is sufficiently strong or the dynamics of neurons are sufficiently slow because the opposite would lead to, in some situations, an antiphase \cite{sherman1992rhythmogenic, cymbalyuk1994phase, han1995dephasing, chow2000dynamics}. However, studies have shown that when the coupling is not too strong, the synchronization of neurons depends on the intrinsic properties of neurons, but also on the combination of electrical synapses with chemical synapses \cite{benjamin2003,siam2020,pre2018sinha,benjamin2005}.

\par Neuronal synchronization being a phenomenon linked to several brain pathologies such as epilepsy, Parkinson diseases, Alzheimer, autism and schizophrenia \cite{benjamin2005}, does not always appear in a singular way in neuronal networks. Its presence (state of coherence) accompanied simultaneously by an asynchronous state (state of incoherence) has been demonstrated by Kuramoto and Battogtokh in the networks of identical oscillators not locally coupled \cite{kuramoto2002}. This phenomenon was later termed the chimera state \cite{abrams2004}. Recently, chimera states have been explored in different systems and using different types of couplings \cite{pr2021,plr2019,epl2017,delay2016,epl2018,neucom2020,epl2020,pre2019}. Chimera states are analogously to the cerebral behaviors of certain aquatic mammals and migratory birds, which during their movements half part of their brains asleep while the rest are awake  \cite{rattenborg2000,rattenborg2006}. The chimera state has several variants \cite{sethia2013,zakharova2014,kapitaniak2014,xie2014,bera2016,abrams2008,li2016} among which we have the traveling chimera which is of particular interest to us in the context of this work.

\par Traveling chimera, for its part, marks a chimera state with displacement in time. That is to say, we are witnessing the presence in the network under study, of patterns of coherent elements interspersed by other incoherent ones in movement and alternately changing their position over time. However, to our knowledge, the existence of this phenomenon has not yet been revealed in a two-dimensional (2D) Hindmarsh-Rose (HR) neuronal network. Unlike chimera states which studies have shown to exist in 1D, 2D and 3D networks  \cite{hizanidis2014,bidesh2016,kundu2018chimera,srilena2019} for some locally coupled and others non-locally coupled, the traveling chimera state was only studied in one-dimensional (1D) networks. In this connection, the work of Hizanidis et al. \cite{hizanidis2015} is inscribed, which highlights the traveling chimera for hierarchical connectivities. Also, Mishra et al. \cite{mishra2017traveling} highlights the traveling chimera in a 1D network of HR neurons coupled locally by an electrical coupling and non-locally by a chemical coupling. Our motivation is to extend the study conducted by Mishra et al. \cite{mishra2017traveling} to a 2D network for the sake of generalization.

\par In this work, we study the mutual influence of electrical and chemical couplings on the dynamics of a 2D neuronal network and verify the appearance of traveling chimera states. For this we first consider the non-local chemical coupling whose variation produces the traveling chimera state in 2D; then we introduce the electrical coupling which, associated with chemical coupling values, for some produces a traveling multi clusters, for others an imperfect traveling chimera, an alternating traveling chimera or even a rotation of the patterns thus changing the direction of propagation. Finally, the energy functions from the Hamiltonian formalism \cite{sarasola2004} are used to deduce the direction of propagation of the traveling chimera in 2D.

\par The rest of the work is presented as follows: In Sec. II, we present the network and its mathematical model.  In Sec. III, we present the phenomenon of traveling chimera state in 2D network. Then the energy analysis is presented to characterize the emerging states in Sec. IV and finally the Sec. V closes with a conclusion and some perspectives.

\section{Model and Network}

\begin{figure}[!h]
\includegraphics[width=7.5cm]{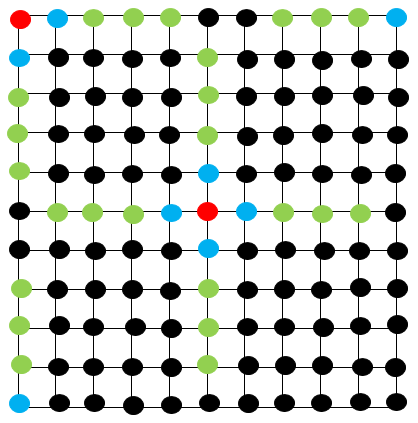}
\caption{\label{fig.sgr1} Schematic diagram of a two-dimensional grid: the $(i,j)$-th node (red circle) is locally connected to four nearest-neighbors (blue circles) and non-locally connected to $p$-nearest-neighbors (green circles). Black circles represent other nodes on the network. For the simple illustration, we choose $M=11$ and $p=4$.}
\end{figure}

We consider a Hindmarsh-Rose neuronal model in the form,
\begin{equation}\label{eq.hr1.mt}
\begin{cases}
\dot{x}=y-ax^3+bx^2-z+I,\\
\dot{y}=1-dx^2-y,\\
\dot{z}=r\left[s\left(x-x_0\right)-z\right].
\end{cases}
\end{equation}
Here,  $x$ represents the membrane potential of the neuron, $y$ is the variable associated with the fast dynamics of the current, $z$ is that associated
with the slow dynamics of the current, $a, b, d, I, r, s$ are the system's parameters whose values are the following: $a=1$, $b=3$, $d=5$, $I=3.5$, $r=0.01$, $s=5$; where $r$
is the time scale parameter which is relatively small, $x_0$ is the first coordinate of the fixed point of the system and $I$ represents
the current applied to the neuron to stimulate it \cite{hr3}. The mathematical equations which characterize the 2D grid of $M \times M$ coupled Hindmarsh-Rose neuron model, is given by the following  of equations,
\begin{equation}\label{eq.hr2.mt}
\begin{cases}
\dot{x}_{i,j}=y_{i,j}-ax_{i,j}^3+bx_{i,j}^2-z_{i,j}+ I +J_{i,j} +C_{i,j},\\
\dot{y}_{i,j}=1-dx_{i,j}^2-y_{i,j},\\
\dot{z}_{i,j}=r\left[s\left(x_{i,j}-x_{i,j0}\right)-z_{i,j}\right],
\end{cases}
\end{equation}
where $J_{i,j}$ represents the nearest neighbor electrical coupling and described as follow,

\begin{equation}\label{eq.hr2_1.mt}
J_{i,j}=\dfrac{k_1}{4}\left(\displaystyle\sum_{l=i-1}^{i+1}\left(x_{l,j}-x_{i,j}\right)+\displaystyle\sum_{l=j-1}^{j+1}\left(x_{i,l}-x_{i,j}\right)\right),
\end{equation}
and $C_{i,j}$ defines the non-local chemical coupling in the form,
\begin{equation}\label{eq.hr2_2.mt}
C_{i,j}=\dfrac{k_2}{4p-4}\left(v_s-x_{i,j}\right)\left(\displaystyle\sum_{\substack{l=i-p \\ p \neq 0, 1}}^{i+p}\Gamma\left(x_{l,j}\right)+\displaystyle\sum_{\substack{l=j-p \\ p \neq 0, 1}}^{j+p}\Gamma\left(x_{i,l}\right) \right),\\
\end{equation}
for $i,j=1,2,...,M$ and $M$ is the total number of elements lying on one side of the 2D grid ($M=100$) with boundary conditions, $x_{0,j}=x_{M,j}$, $x_{i,0}=x_{i,M}$, $x_{M+1,j}$ $= x_{1,j}$ and $x_{i,M+1}$ $= x_{i,1}$, $v_s$ is the reversal potential that determine the connection to be inhibitory or excitable depending on whether $v_s$ is greater or less than $x_{i,j}$. We chose $v_s=2.0$ so that the neurons interact through excitatory chemical synapses. $k_1$ and $k_2$ are respectively the coefficients of gap junctional and  chemical synaptic coupling. An element of the 2D network is identified by its coordinates $(i, j)$. p is the number of neighbors to the left, to the right, to the bottom and to the top non-locally coupled to the element of position $(i, j)$ of the grid via chemical synapses ($p=40$); with the exception of the four first nearest  neighbors which are rather locally coupled to the same element through the electrical synapses. A local illustration of the 2D network forming a lattice square is shown in \ref{fig.sgr1} and illustrates the connectivity of two nodes (in red): one in the middle of the grid and the other at the border. $\Gamma\left(x_{i,j}\right)$ represents the chemical synaptic coupling function. It is modeled by the sigmoidal function in the form,
\begin{equation}
	\label{eq.gamma.mt}\Gamma\left(x_{i,j}\right)=\dfrac{1}{1+\exp\big (-\lambda(x_{i,j}-\theta_s)\big )},
\end{equation}
where $\lambda=10$ is the slope of sigmoidal function, $ \theta_s=-0.25$ is the synaptic firing threshold. The initial conditions of the network are given by: $x_{i,j}(0)=0.001(M-(i+j))+\zeta_{xij}, y_{i,j}(0)=0.002(M-(i+j))+\zeta_{yij}, z_{i,j}(0)=0.003(M-(i+j))+\zeta_{zij}$, where $\zeta_{xij}$, $\zeta_{yij}$, $\zeta_{zij}$ are small random fluctuations.

\section{Traveling chimera in two-dimensional network}

\begin{figure}[!h]
	\includegraphics[width=8.5cm]{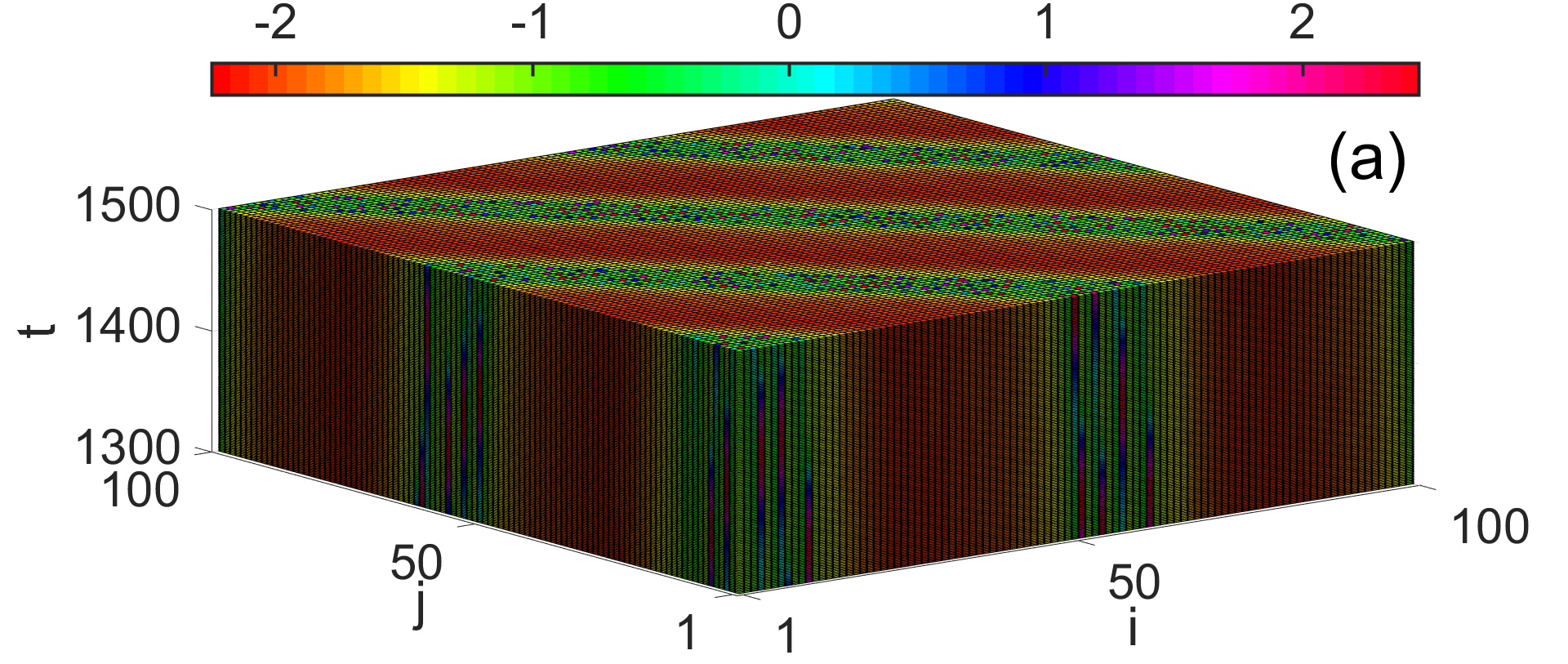}
	\includegraphics[width=8.5cm]{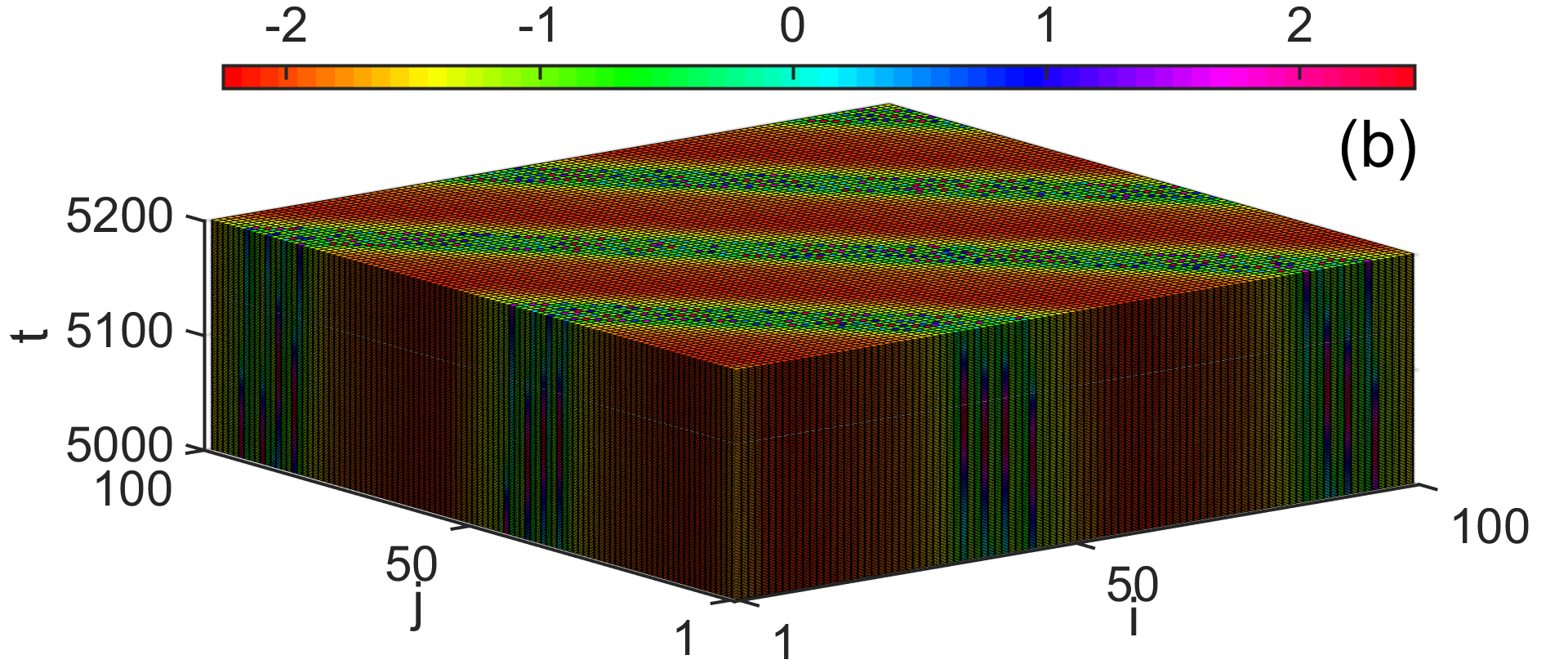}
	\includegraphics[width=8.5cm]{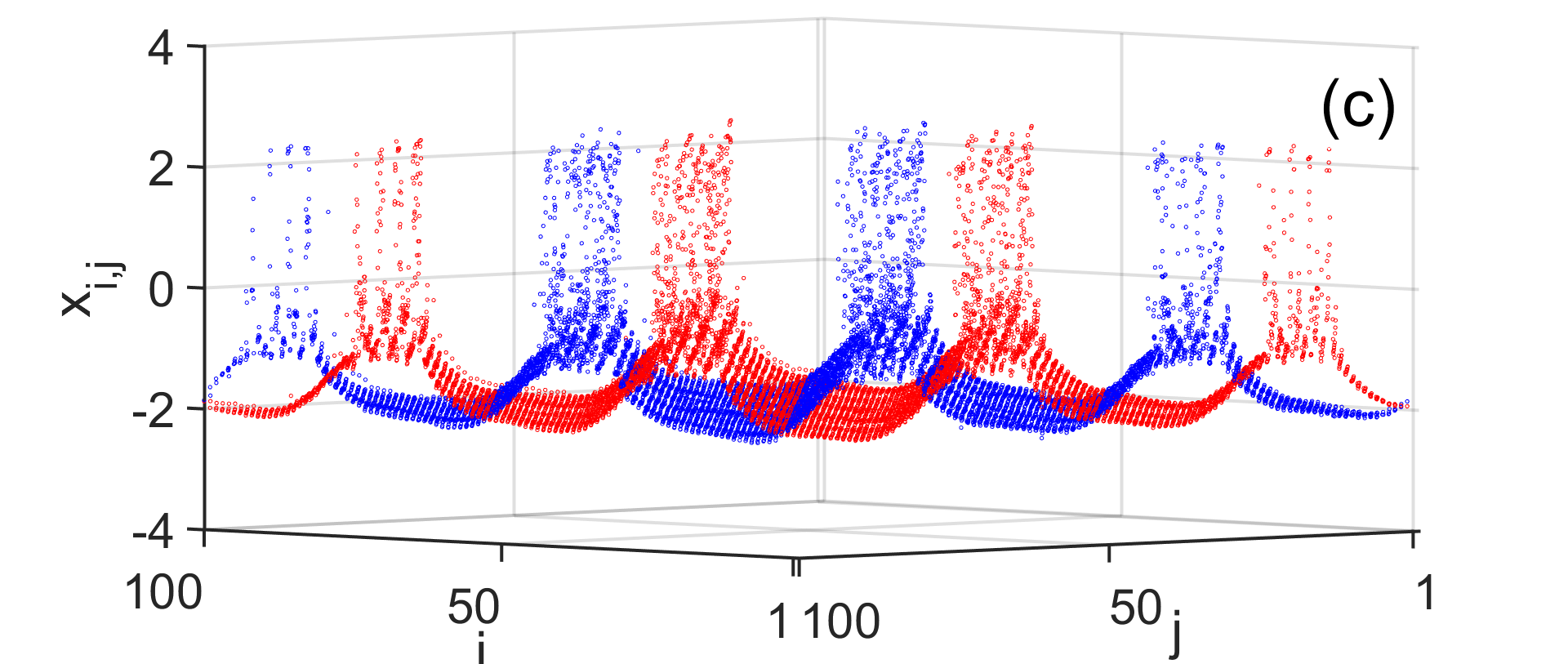}
	\caption{\label{fig.sgr2}Traveling chimera patterns in 2D network for synaptic coupling $k_2=9$ ($k_1=0$). (a,b) Spatiotemporal evolution of $x_{i,j}$ for $M^{2}$ neurons, (c) snapshots of $M^{2}$ number of $x_{i,j}$ variables in 2D plane at two different instant of times. Blue curve is for a later time than the red snapshot implying a traveling pattern. The color bars in (a, b) represent the variation of $x_{i, j}$.}
\end{figure}

We perform a numerical study of the elements of the 2D network described by the system of equations (2). The integration of the elements of the network is made through the 5th order Runge-Kutta method. It has being established that in the presence of the local electrical coupling only, the whole network presents only two possible states, {\it i.e.,} coherent and incoherent states \cite{kundu2018chimera}. For numerical simulations, we first consider the chemical synaptic coupling and vary the corresponding coupling strength $k_2$. The passage of the chemical coupling strength to relatively large values ($k_2$ = 9) allows us to observe that in the 2D lattice, there appear transverse groups of elements which are coherent and which alternate with other incoherent ones. Observation of the 2D network over time clearly shows that these two sets move periodically and alternately as shown in Figs. \ref{fig.sgr2}a and \ref{fig.sgr2}b. We represent on two time ranges of the same width (1300 to 1500 and from 5000 to 5200) a 3D section of amplitudes evolution of the elements in the network over time. It does appear a clear shift of the different patterns. This description is further confirmed by Fig. \ref{fig.sgr2}c which shows the snapshots of each neuron's amplitudes taken at two different instants of time differentiated by the colors (respectively blue and red). We can see that there is a shift between these two curves. Hence the notion of traveling chimera, because qualitatively, it appears on the 2D network, two types of areas including coherent zones and others incoherent.

\begin{figure}[!h]
	\includegraphics[width=8.5cm]{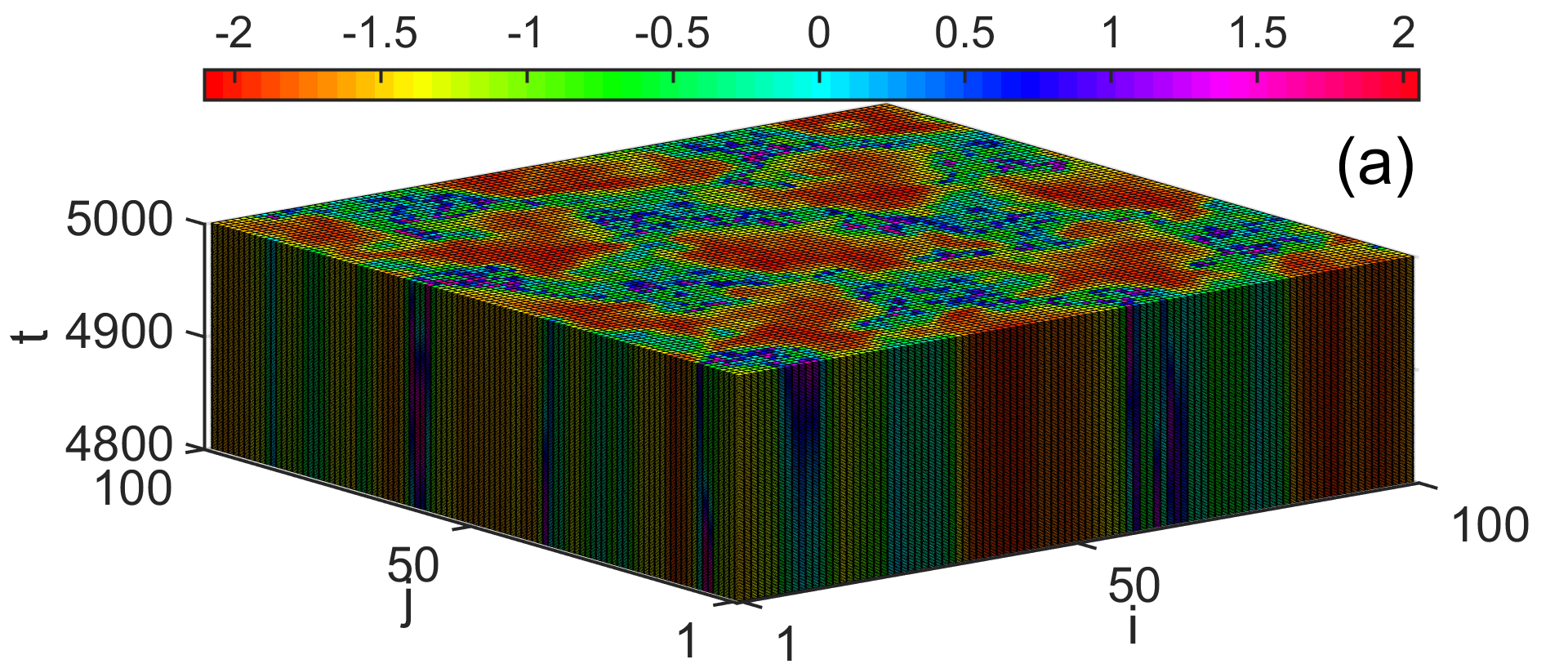}
	\includegraphics[width=8.5cm]{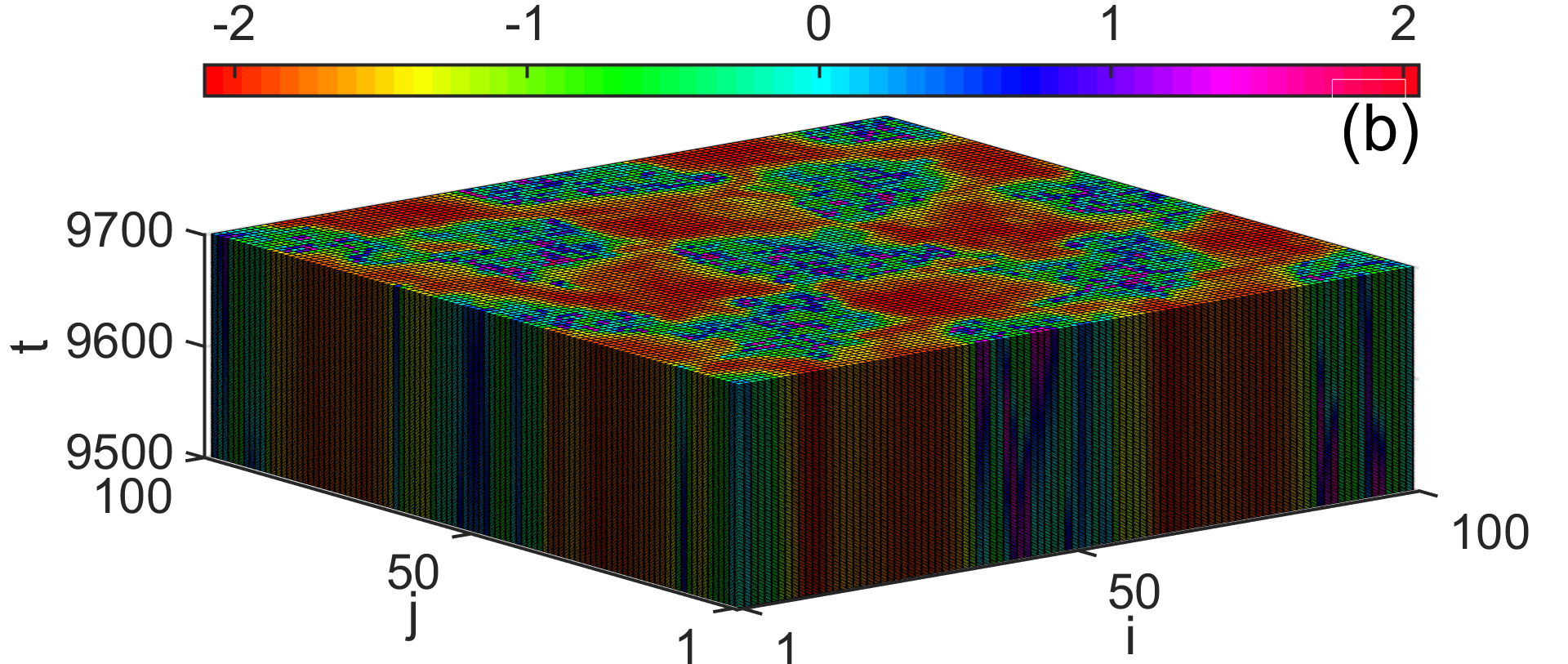}
	\caption{\label{fig.sgr3} Traveling multi-cluster in 2D network for electrical coupling $k_1=1$ and synaptic coupling $k_2=1$. (a,b) Spatiotemporal evolution of $x_{i,j}$ for $M^{2}$ neurons at two different times.}
\end{figure}

Now we consider the contribution of electrical coupling to this dynamic. First, we assign the two coupling forces values $k_1$ and $k_2$ respectively equal to unity. It appears a degeneration of the traveling chimera which produces a reorganization of all the elements of the grid into several groups of coherent elements separated from each other by other groups of incoherent elements as shown in Fig. \ref{fig.sgr3}(a). These coherent groups of elements thus form 2D clusters. Moreover, the observation over time of the 2D evolutions of coherent groups shows that the latter are not static. They move and change positions with slight modifications to their overall 2D structures (spatial configuration) which can be seen comparing Figs. \ref{fig.sgr3}(a) and \ref{fig.sgr3}(b). This ability of coherent blocks to move, especially in time, gives it the qualifier of Traveling multi-clusters in 2D network.

\begin{figure}[!h]
	\includegraphics[width=8.5cm]{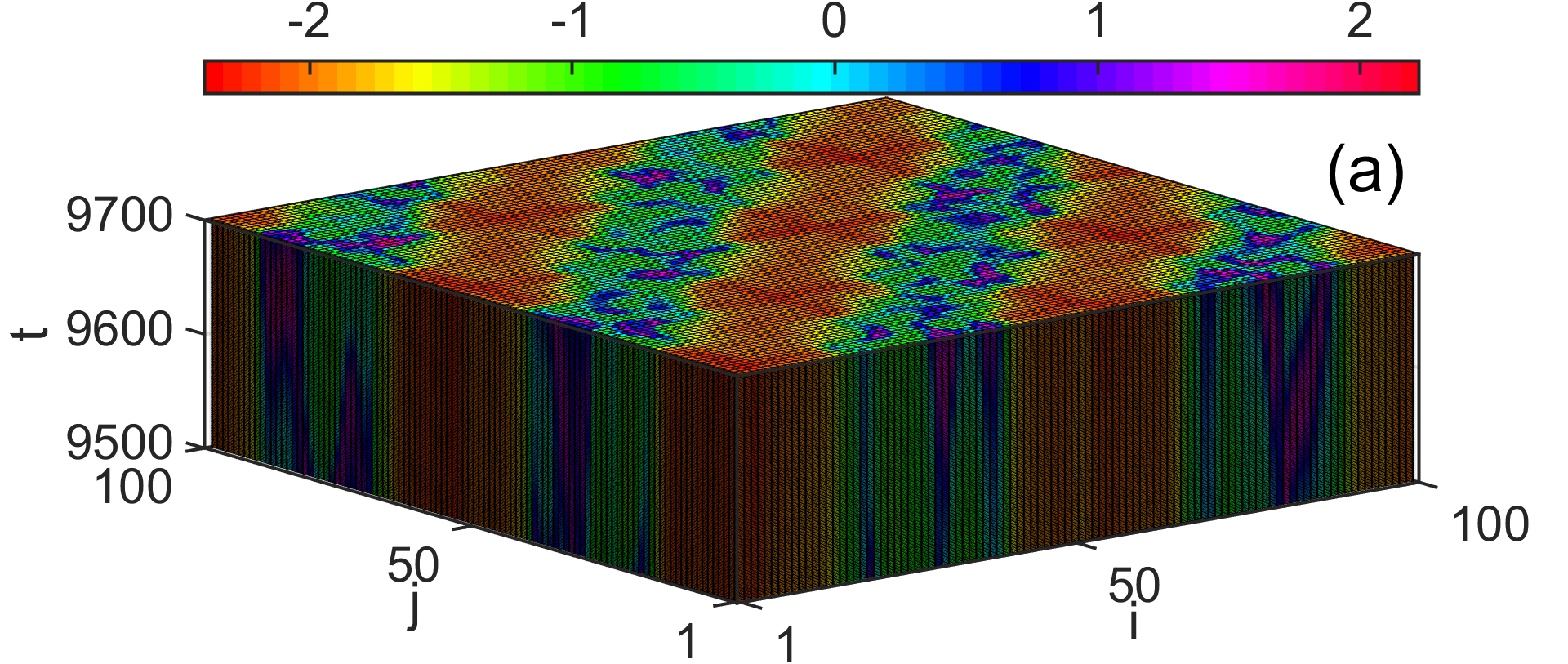}
	\includegraphics[width=8.5cm]{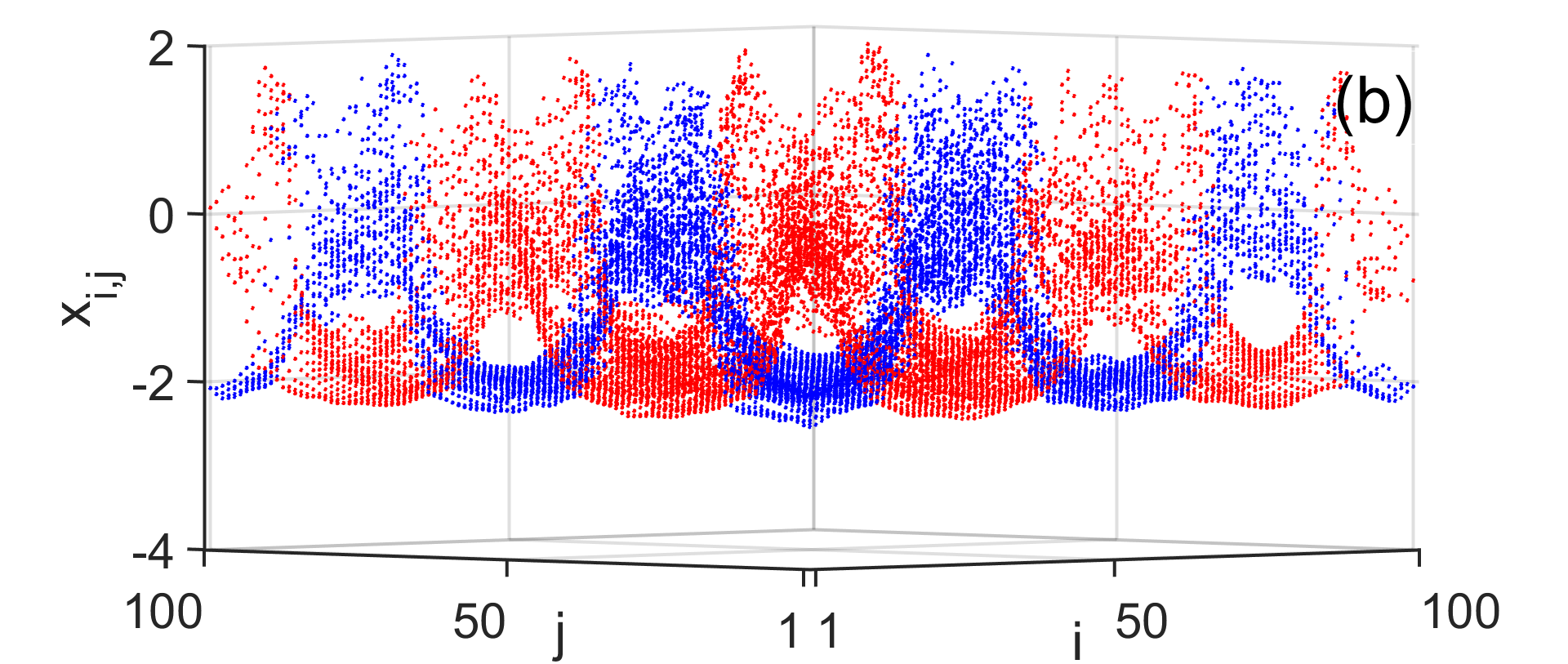}
	\caption{\label{fig.sgr4}Traveling chimera patterns in 2D network for synaptic coupling $k_2=2$ and electrical coupling $k_1=3$.(a) Spatiotemporal evolution of $x_{i,j}$ for $M^{2}$ neurons, (b)snapshots of the $M^{2}$ $x_{i,j}$ variables in two differents instant of times. Blue curve is for a later time than the red snapshot implying a traveling pattern.}
\end{figure}

  Now, varying the values of $k_1$ and $k_2$ we notice that when  the value of the chemical coupling strength constant is ($k_2=3$) and vary that of the electrical coupling strength ($k_1=2$),there is a return to the initial overall dynamics: coherent transverse patterns are formed again, alternating with other incoherent patterns (Fig. \ref{fig.sgr4}). In addition, these patterns periodically move in space. Unlike the case in the absence of electrical coupling, it appears that both coherent and incoherent bands do not have rectilinear shaped boundaries. Also, we notice the change in the direction of propagation of the patterns. This  can be seen by the change in the orientation of the patterns. Instead of having bands parallel to the second diagonal of the large square formed by the lattice, we have bands parallel to the first diagonal. This suggests that the angle of rotation for the direction of propagation of the patterns is around $90^{\circ}$.

\begin{figure}[!h]
	\includegraphics[width=8.5cm]{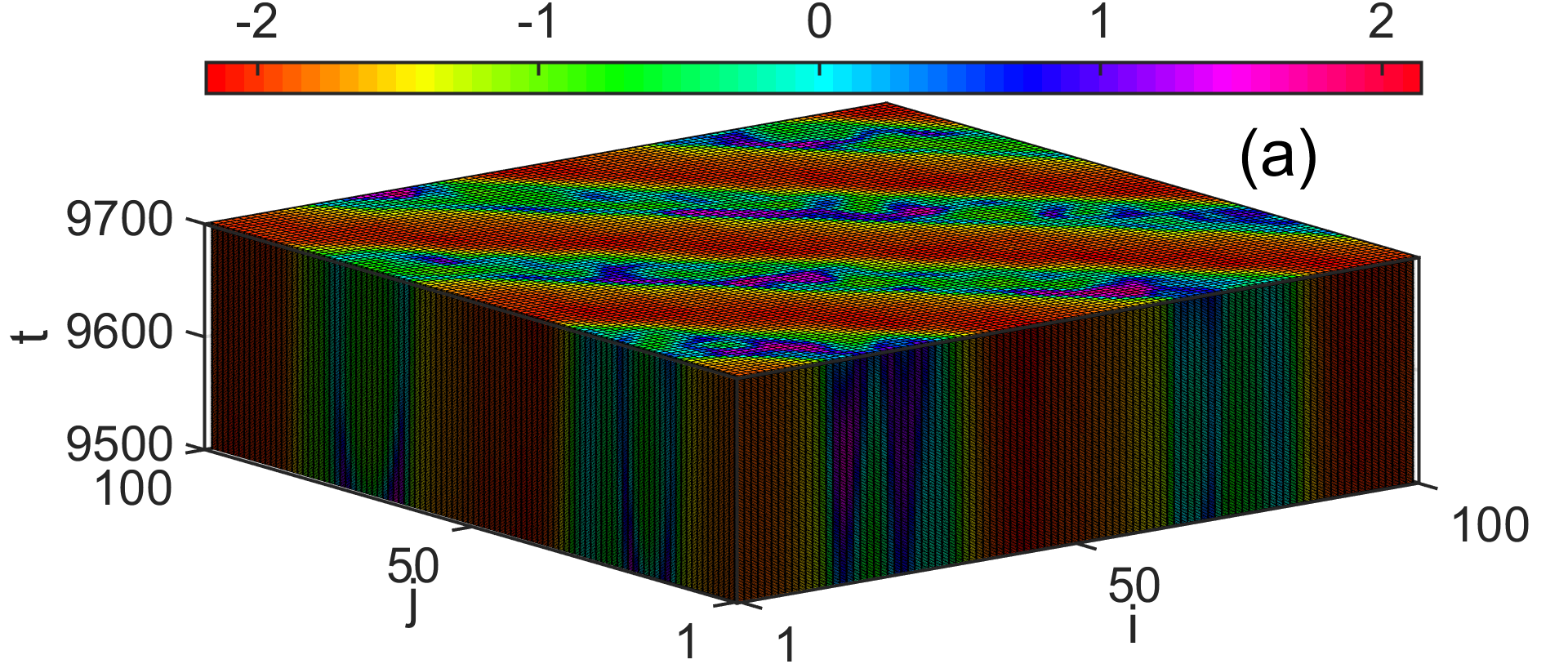}
	\includegraphics[width=8.5cm]{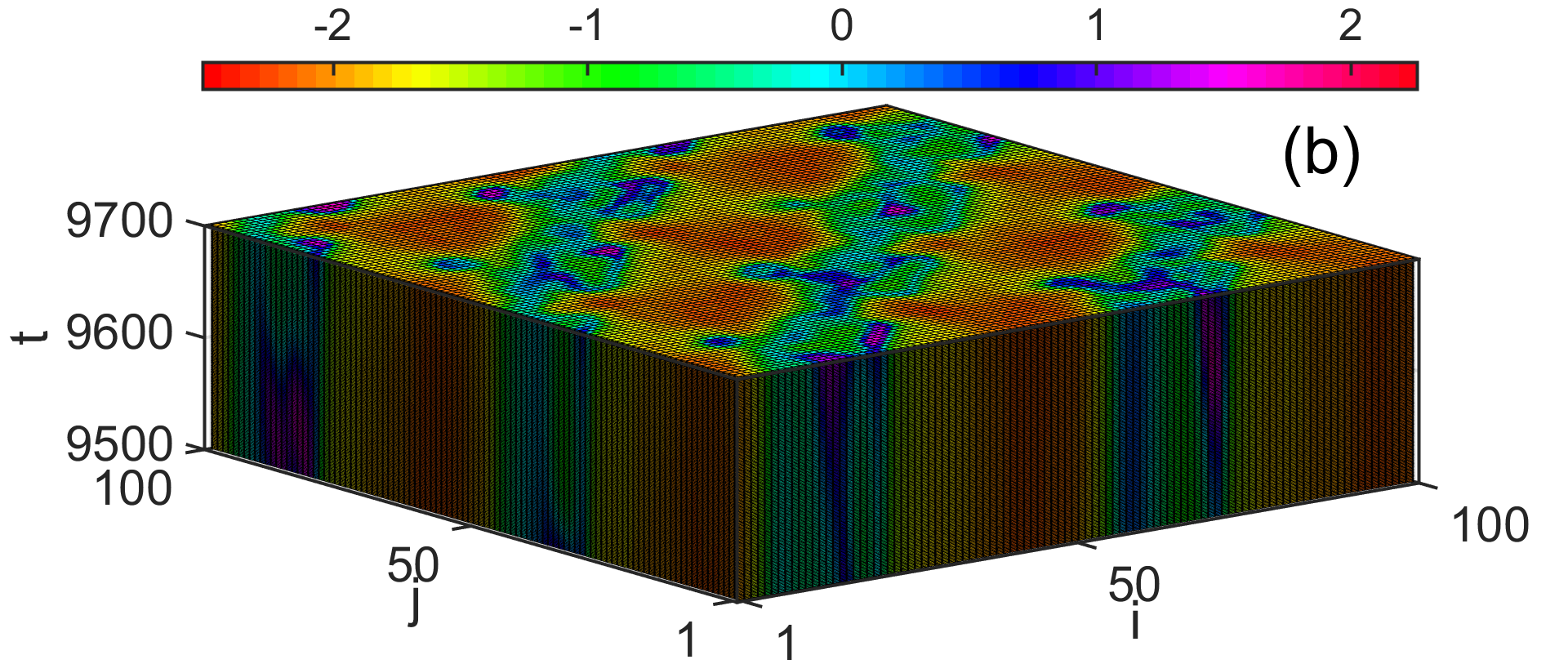}
 	\includegraphics[width=8.5cm]{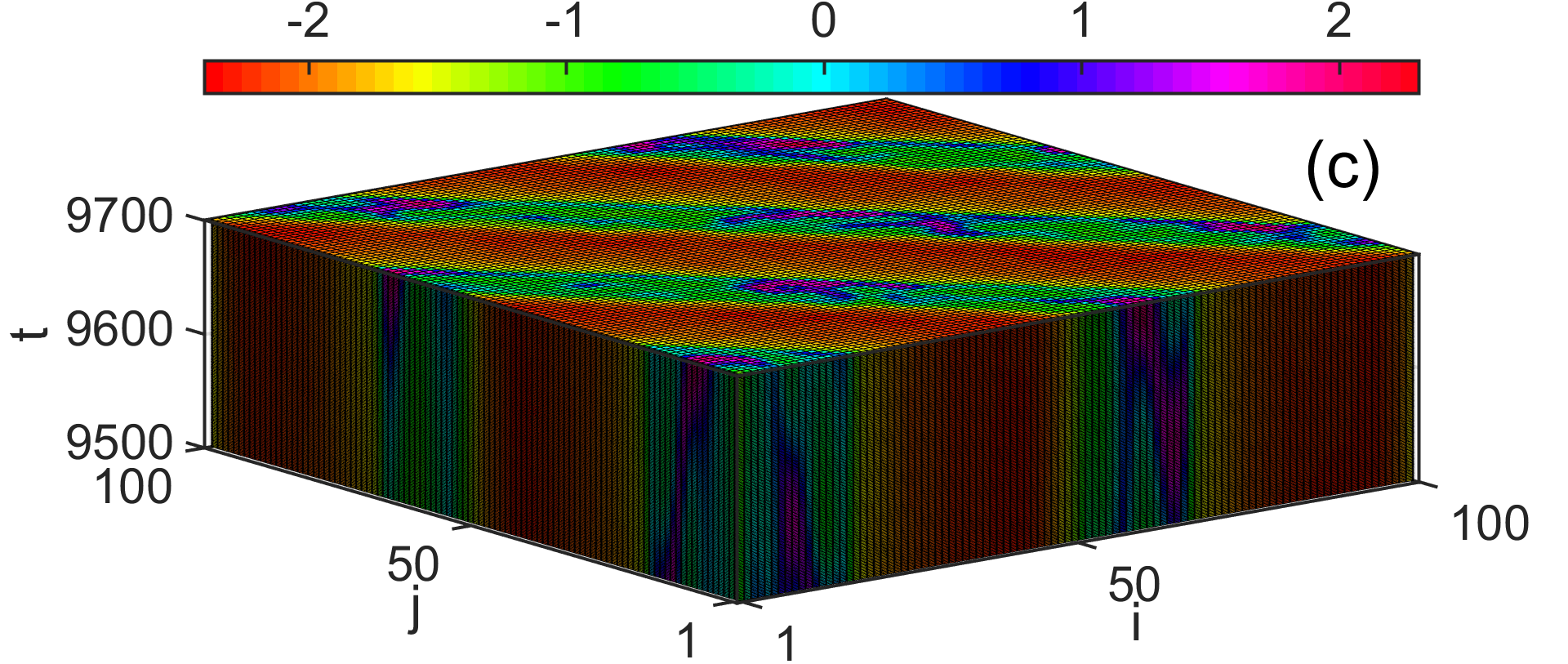}
    \caption{\label{fig.sgr5}Traveling chimera patterns in 2D network for synaptic and electrical coupling with $k_1=5$, (a) $k_2=2.5$, traveling chimera state, (b) $k_2=2.75$, imperfect traveling chimera patterns, (c) $k_2=3$, traveling chimera state. Here, the direction of traveling pattern in (b) changes with the direction of (a, c).}
\end{figure}

We assign the electric coupling strength at the value $k_1=5$ and vary the chemical coupling strength $k_2$ by assigning it successively the values 2.5, 2.75 and 3. It emerges that for the first and the third value ($k_2$ = 2.5, 3), the traveling chimera is established in an identical way to the case in the absence of the electrical coupling, magenta as in Fig.\ref{fig.sgr2}. In fact, the coherent and incoherent bands for the values in these cases are parallel to the second diagonal of the large square formed by the lattice (Figs. \ref{fig.sgr5}(a) and \ref{fig.sgr5}(c)). However, the attribution to $k_2$ of the value 2.75 changes the direction of propagation of the traveling chimera, {\it i.e.,}  the coherent and incoherent bands are parallel to the first diagonal of the large square formed by the network (Fig. \ref{fig.sgr5}(b)). Also, the traveling becomes imperfect.

\begin{figure}[h!]
	\begin{minipage}[c]{.46\linewidth}
		\includegraphics[height=2.6cm,width=4.5cm]{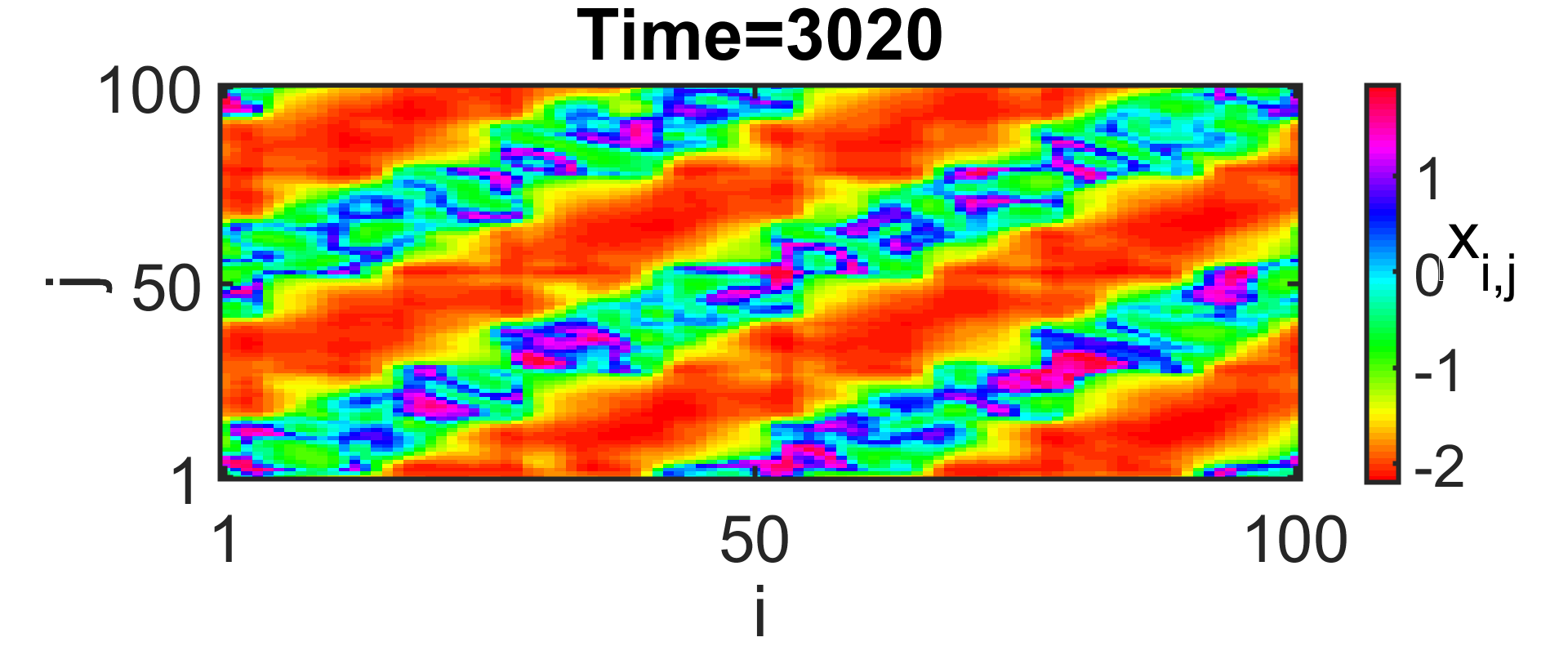}
		\includegraphics[height=2.6cm,width=4.5cm]{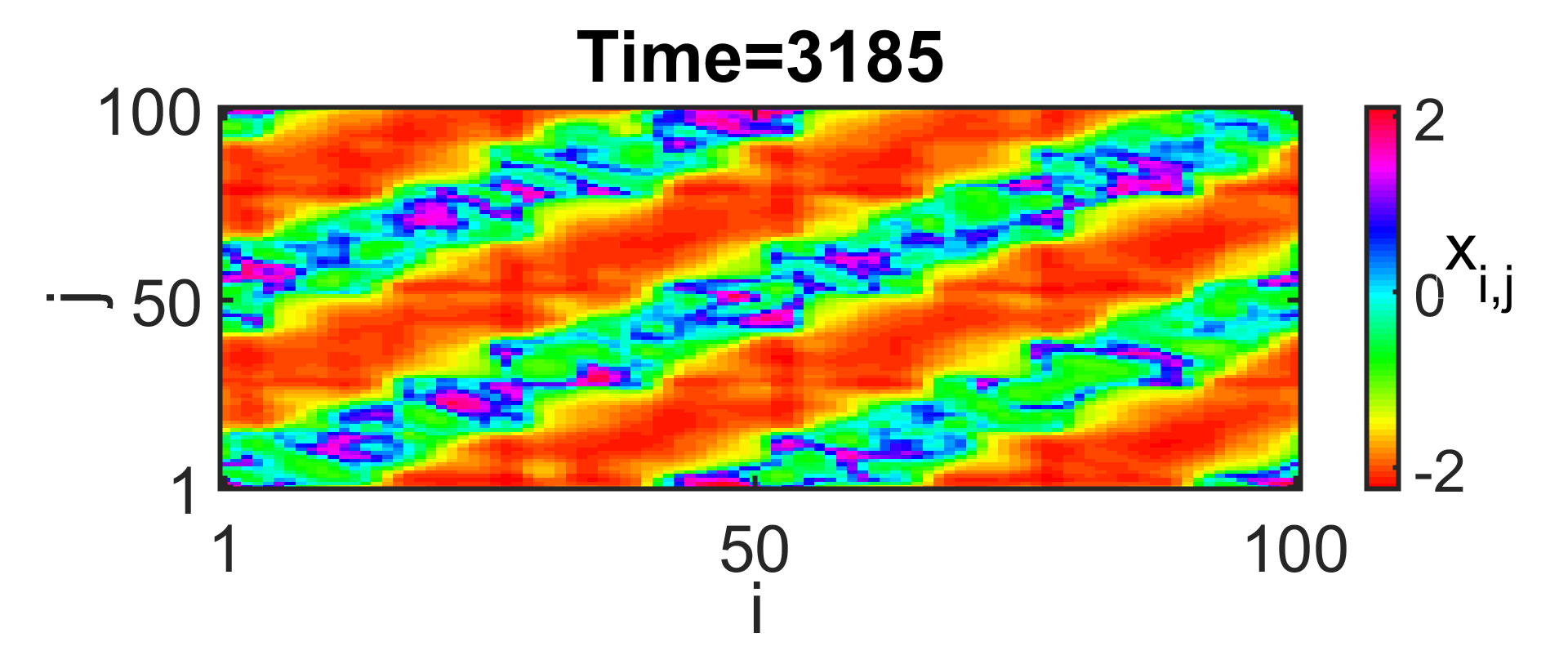}
		\includegraphics[height=2.6cm,width=4.5cm]{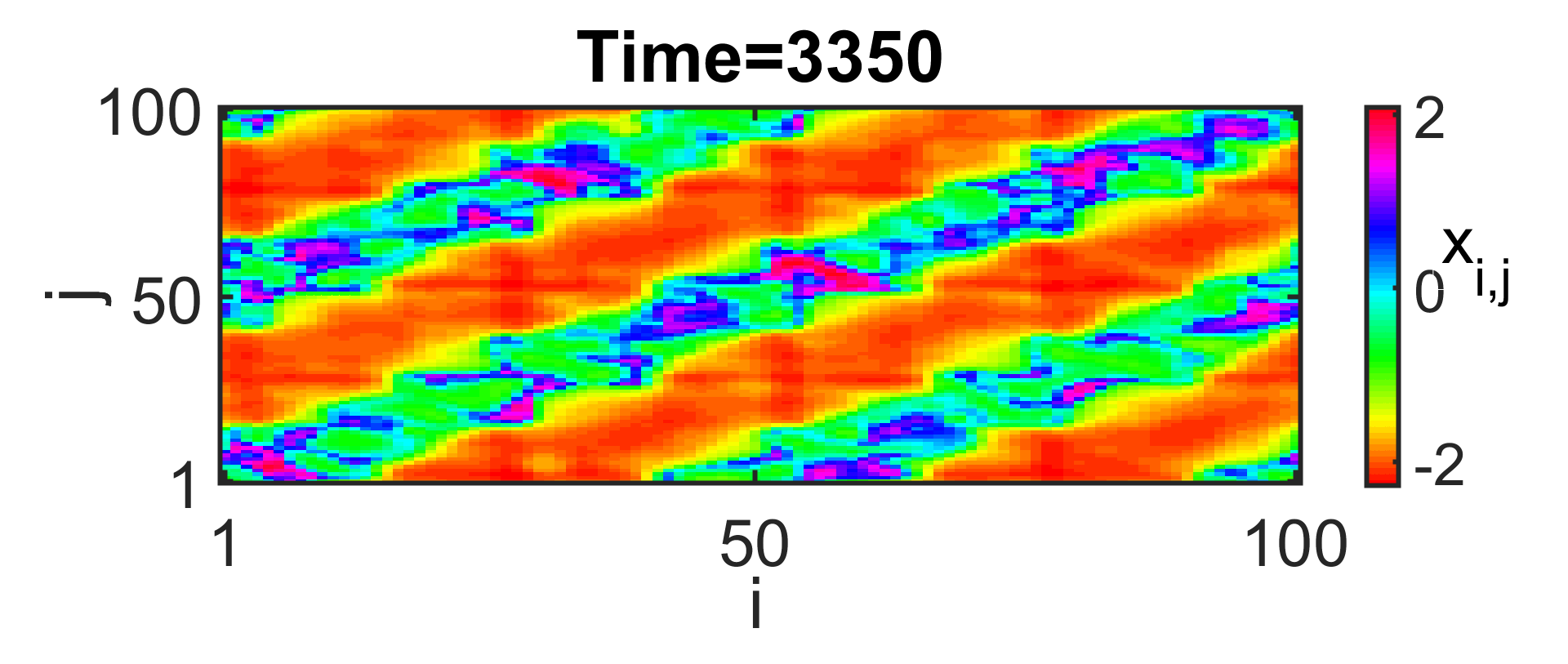}
		\includegraphics[height=2.6cm,width=4.5cm]{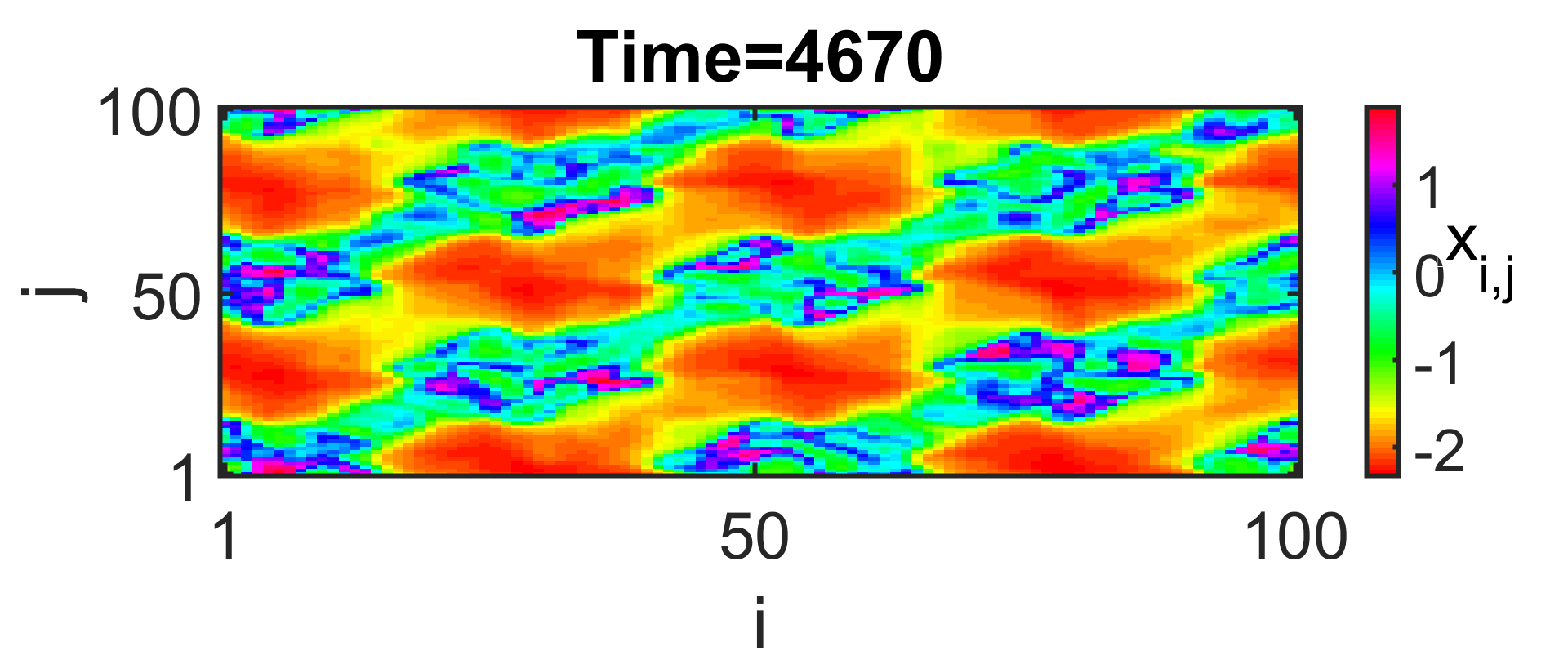}
		\includegraphics[height=2.6cm,width=4.5cm]{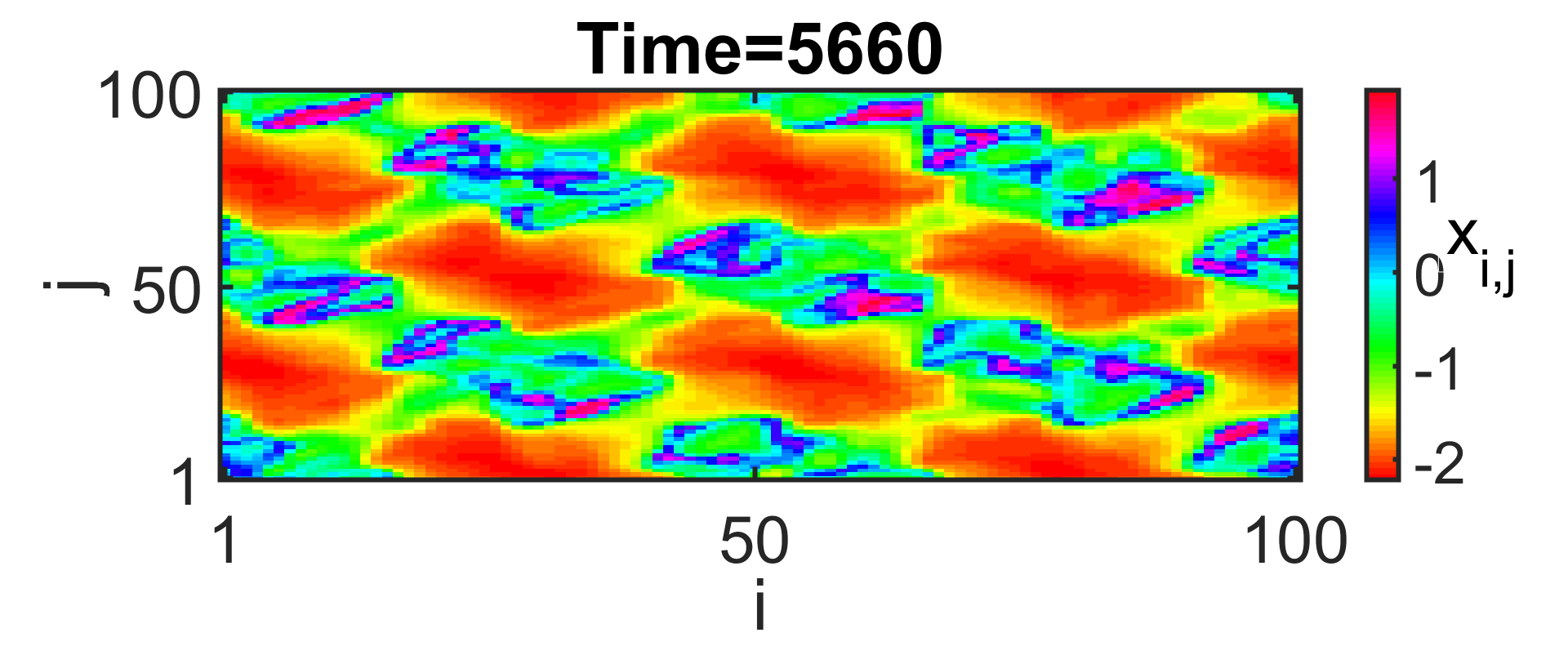}
		\includegraphics[height=2.6cm,width=4.5cm]{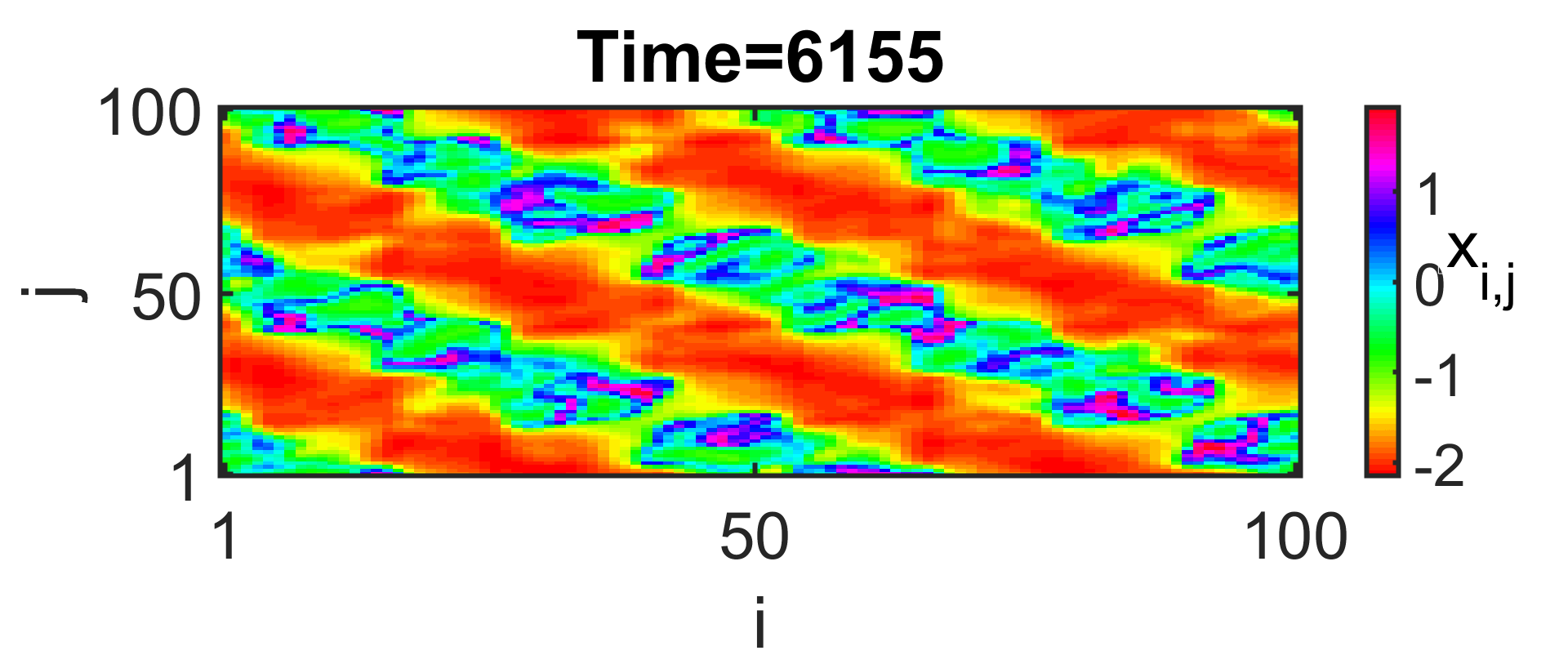}
	\end{minipage} \hfill
	\begin{minipage}[c]{.46\linewidth}
		\includegraphics[height=2.6cm,width=4.5cm]{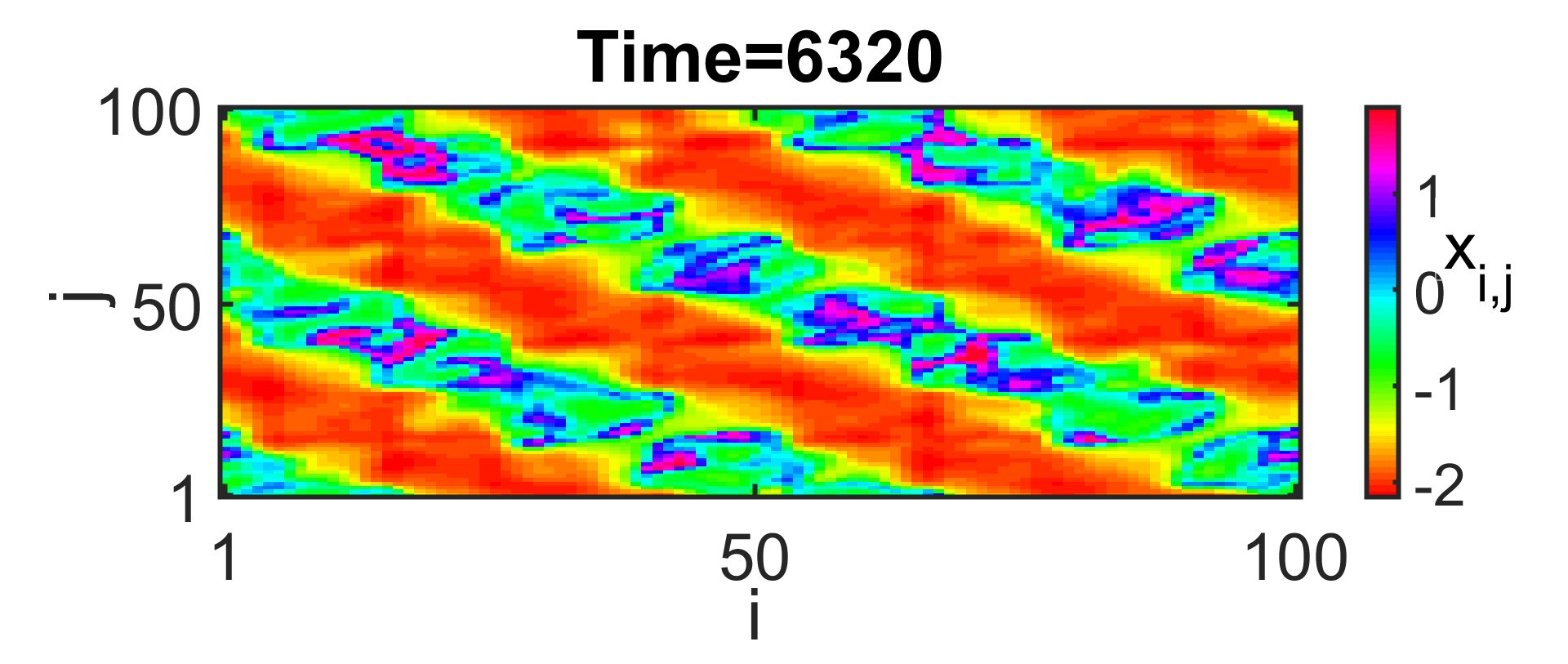}
		\includegraphics[height=2.6cm,width=4.5cm]{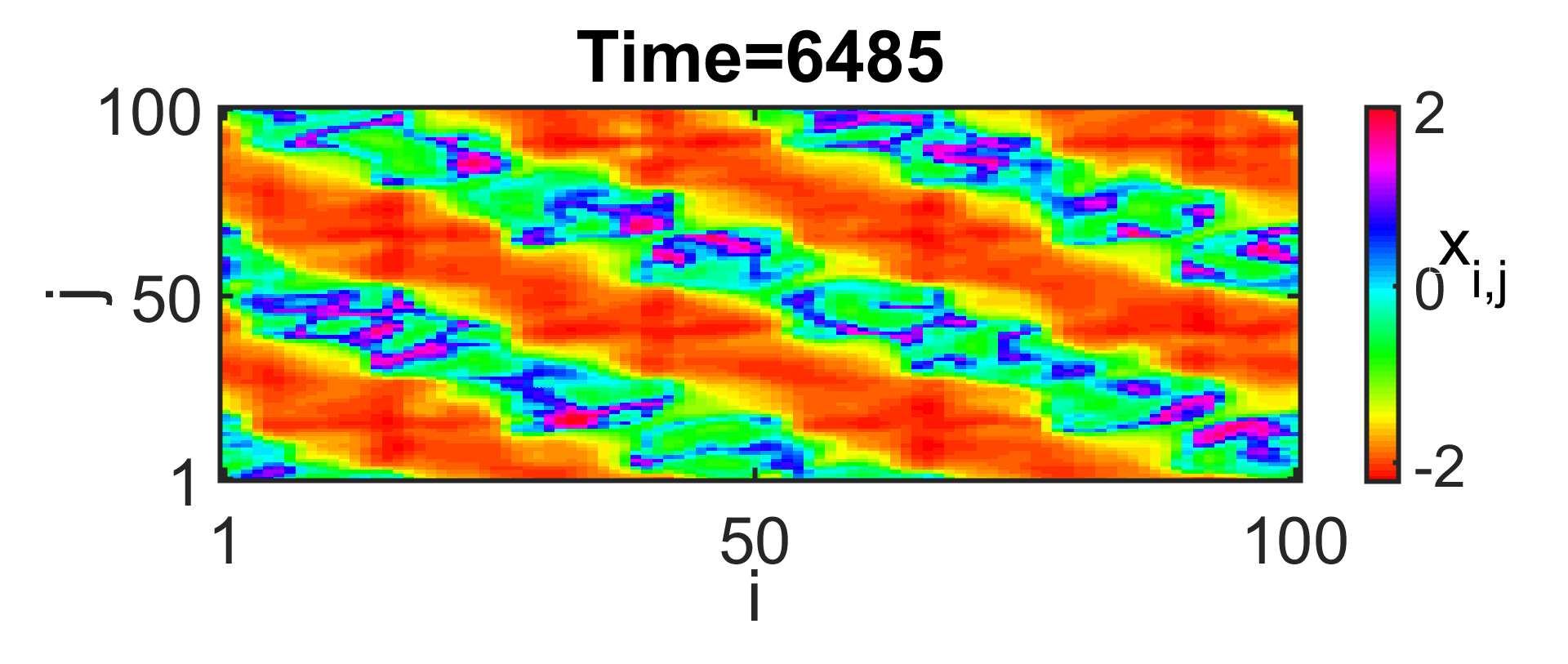}
		\includegraphics[height=2.6cm,width=4.5cm]{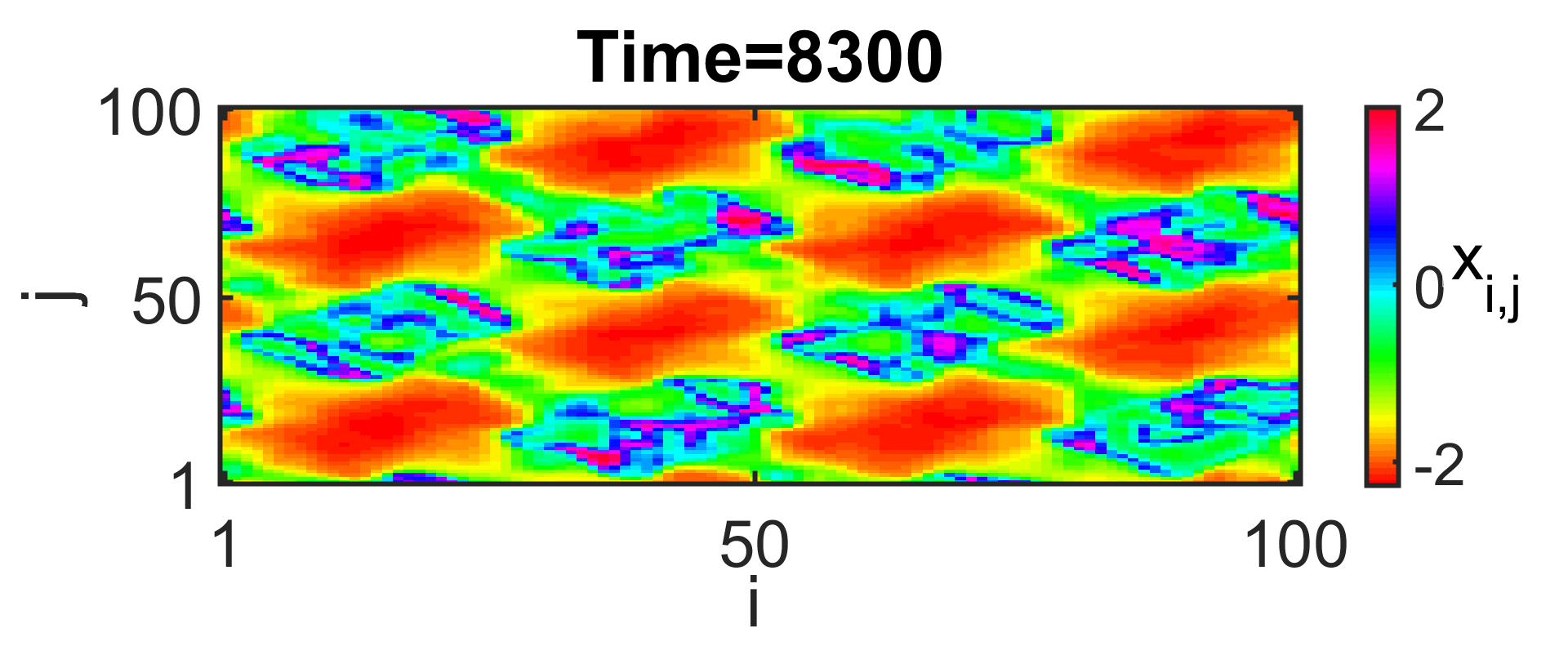}
		\includegraphics[height=2.6cm,width=4.5cm]{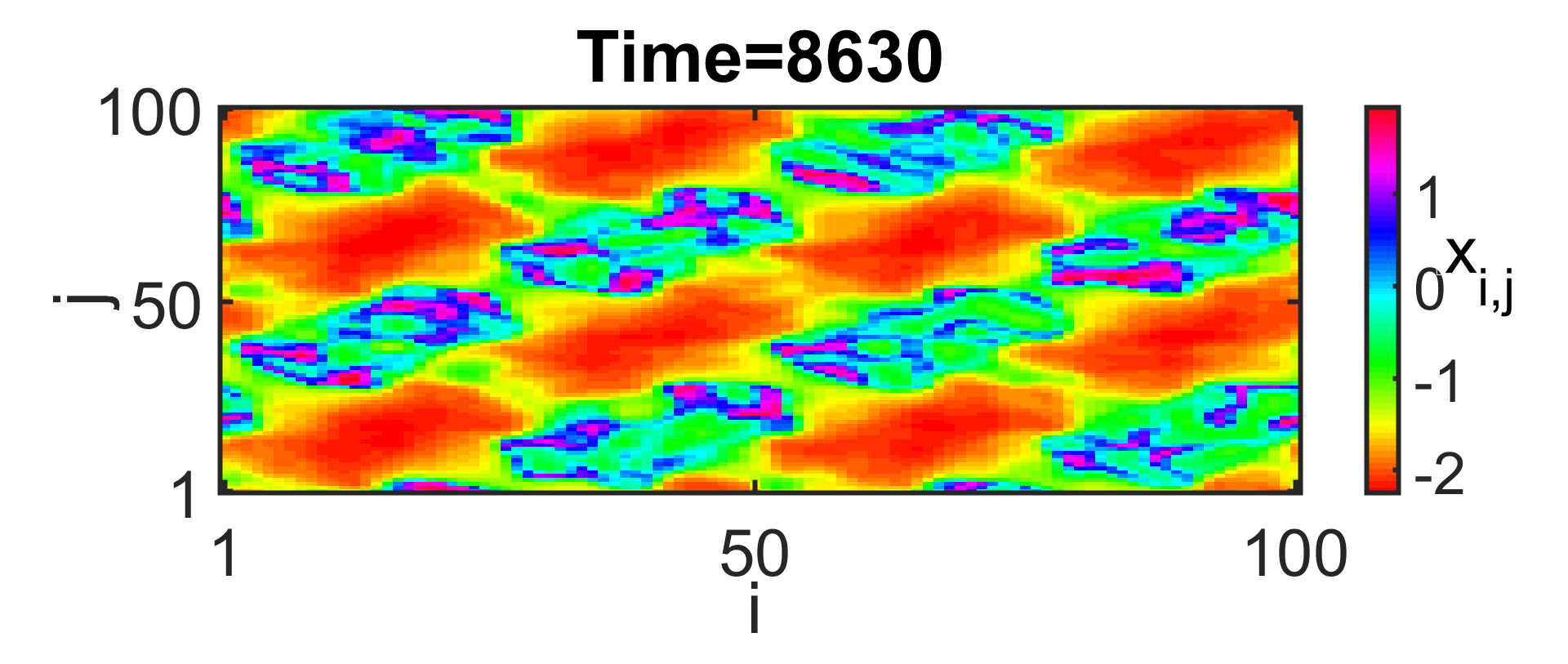}
		\includegraphics[height=2.6cm,width=4.5cm]{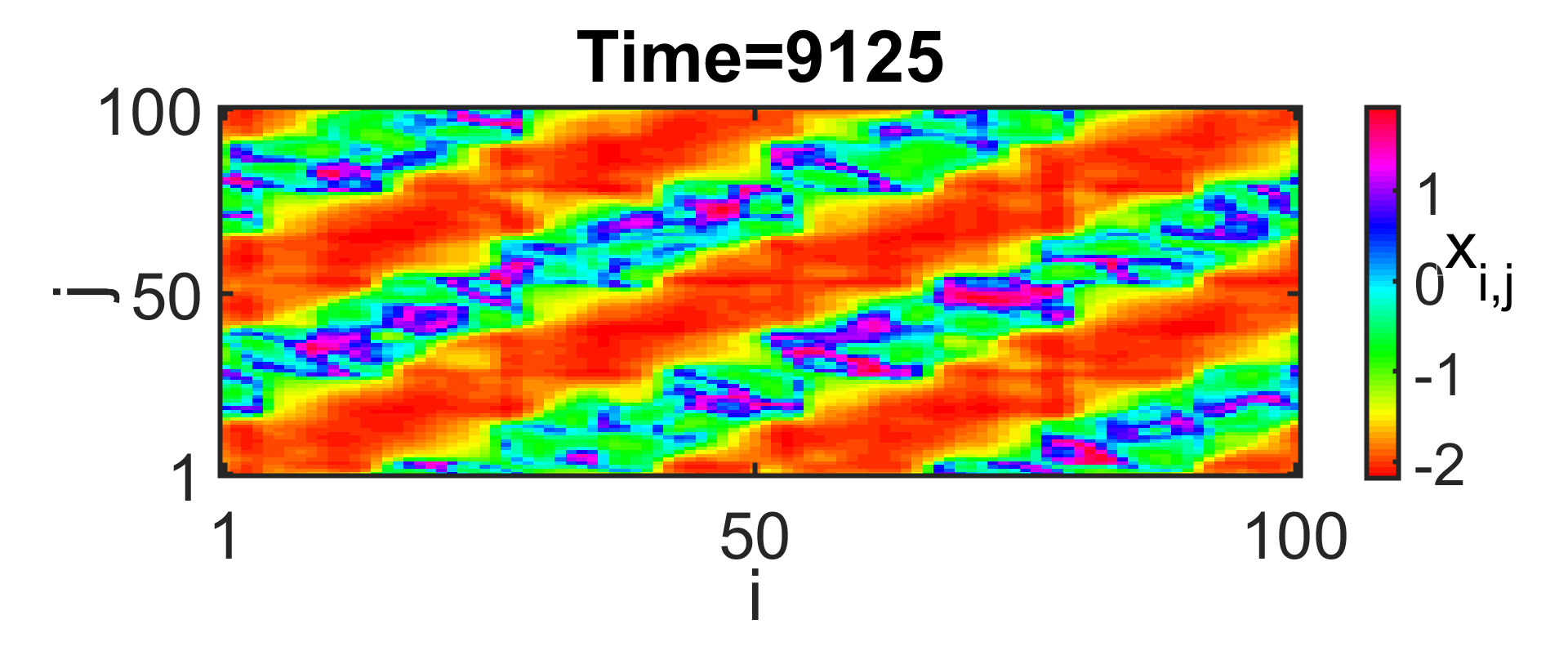}
		\includegraphics[height=2.6cm,width=4.5cm]{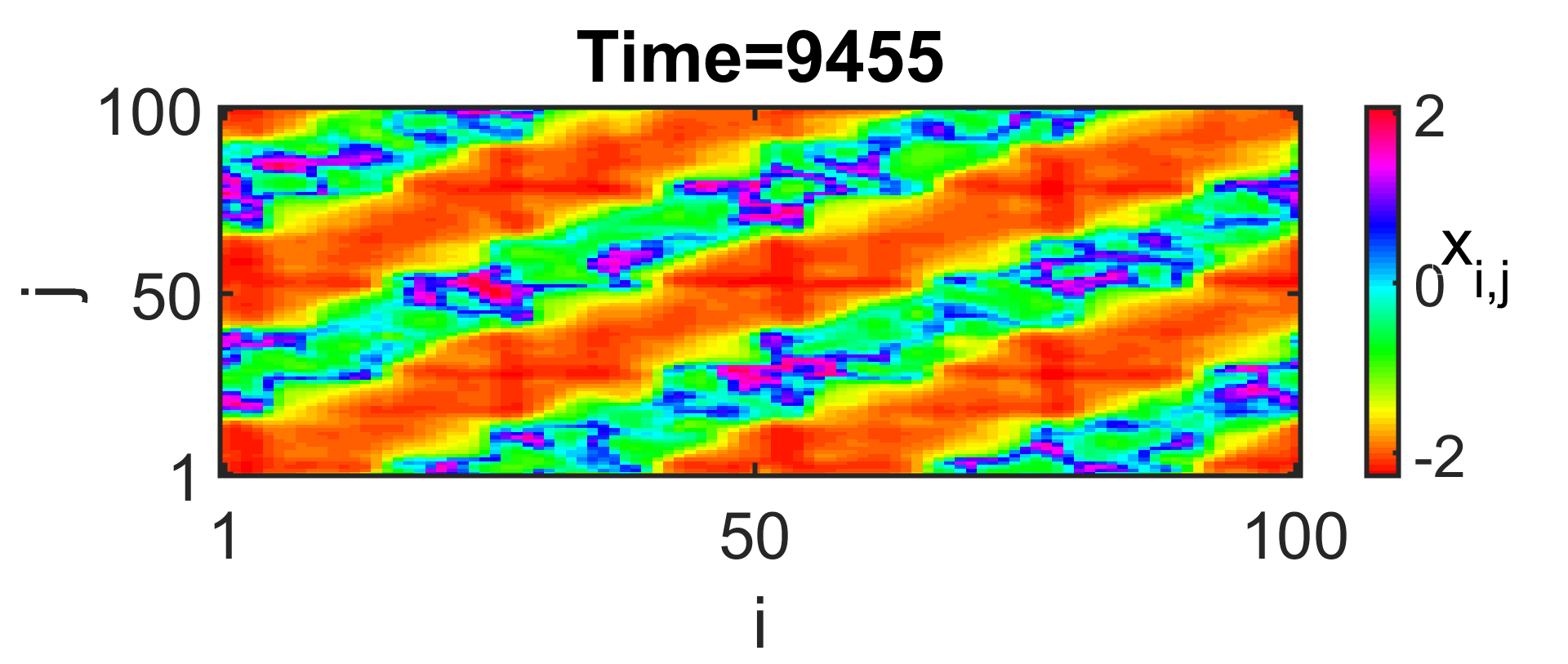}	
	\end{minipage}
	
	\caption{\label{fig.sgr6} 
		Alternating traveling chimera state in 2D $(i, j)$ plane for $k_1 = 3$ and $k_2 = 1$. At time $t$=3020, 3185, 3350 pattern orientation from left to right; time $t$=4670, 5660, 6155 transition for changing orientation; time $t$=6320, 6485 patterns orientation from right to left. With the same repetition thereafter. The color bars represent the variation of $x_{i, j}$. }
\end{figure}

\begin{figure}[!h]
	\includegraphics[height=4.5cm,width=8.5cm]{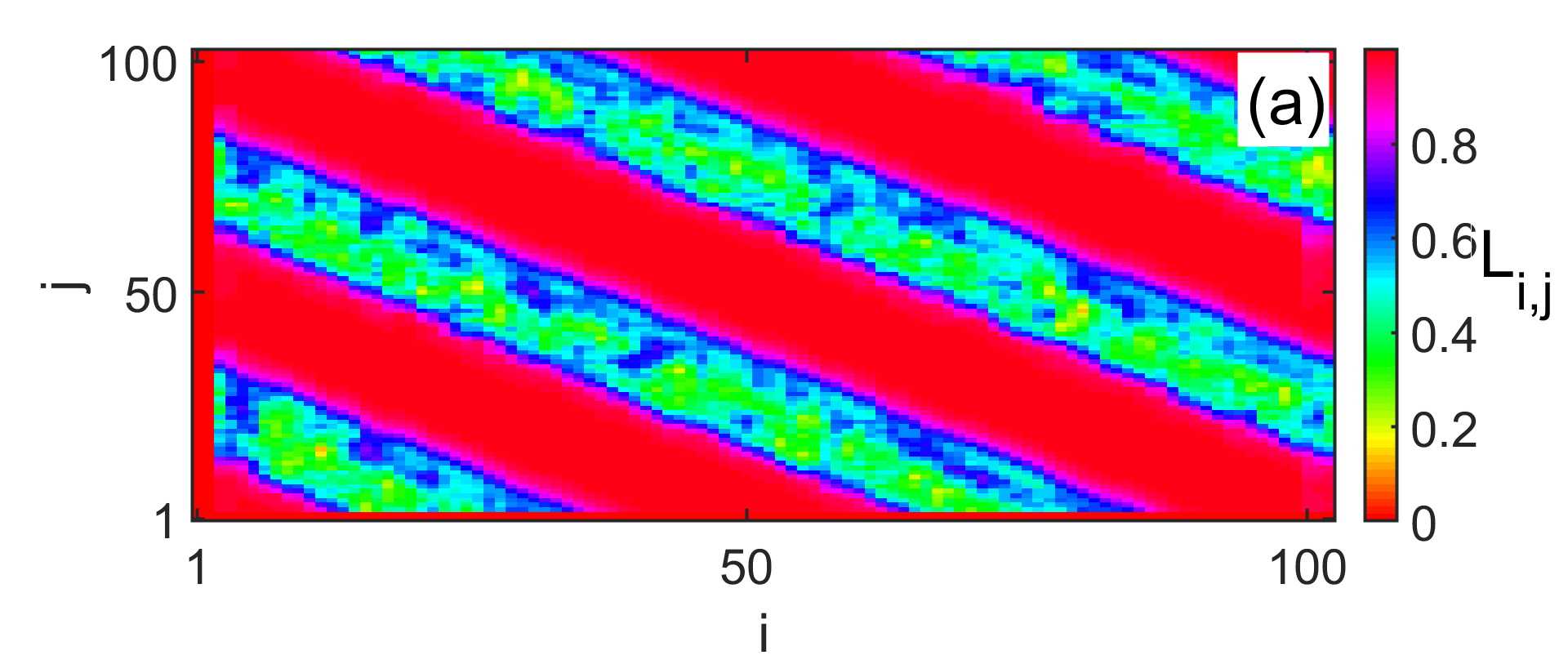}
	\includegraphics[height=4.5cm,width=8.5cm]{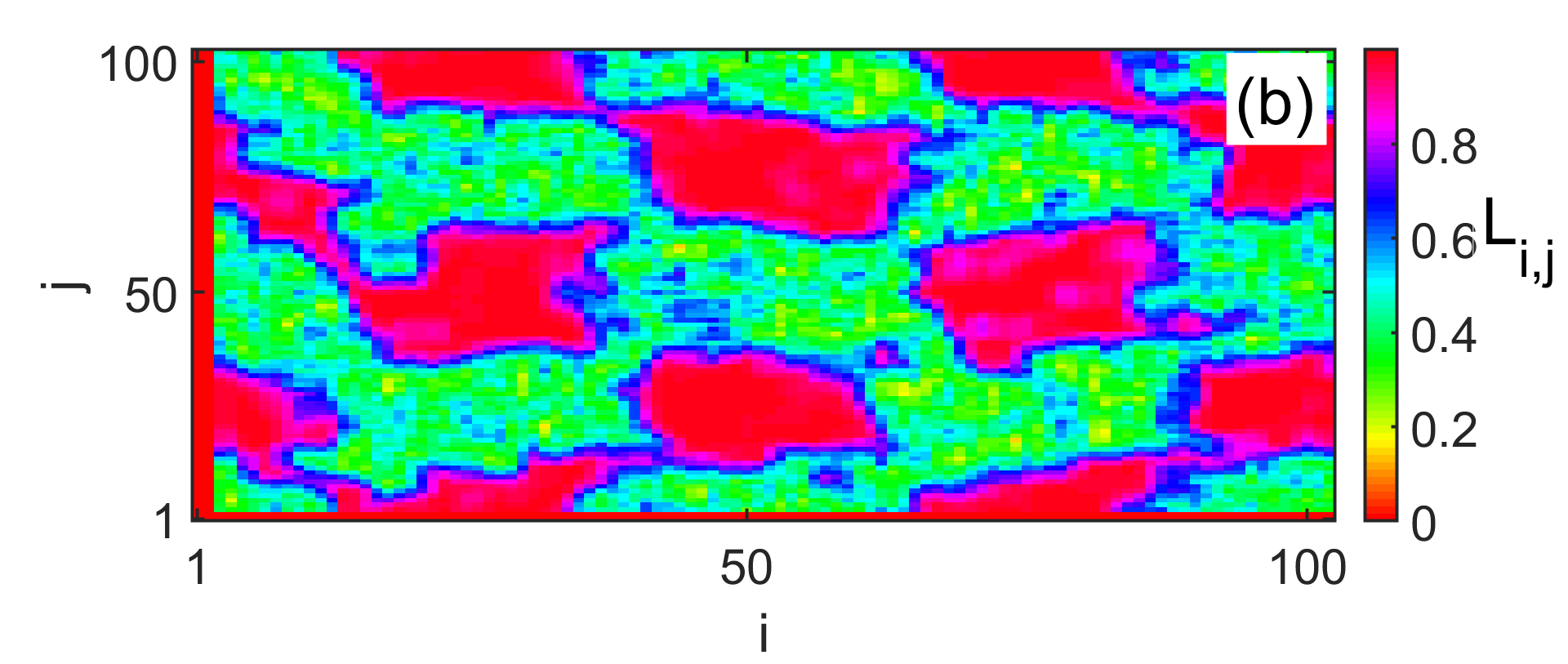}
	\includegraphics[height=4.5cm,width=8.5cm]{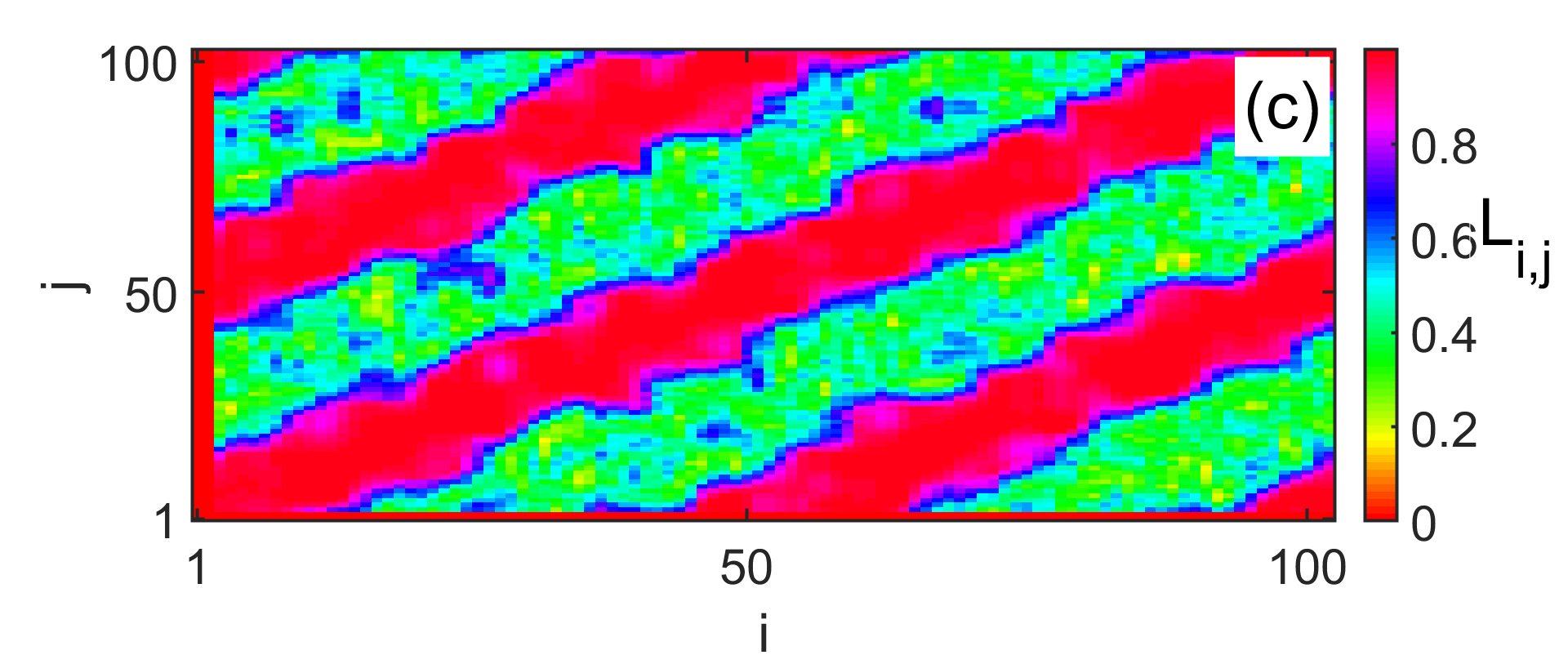}
	\includegraphics[height=4.5cm,width=8.5cm]{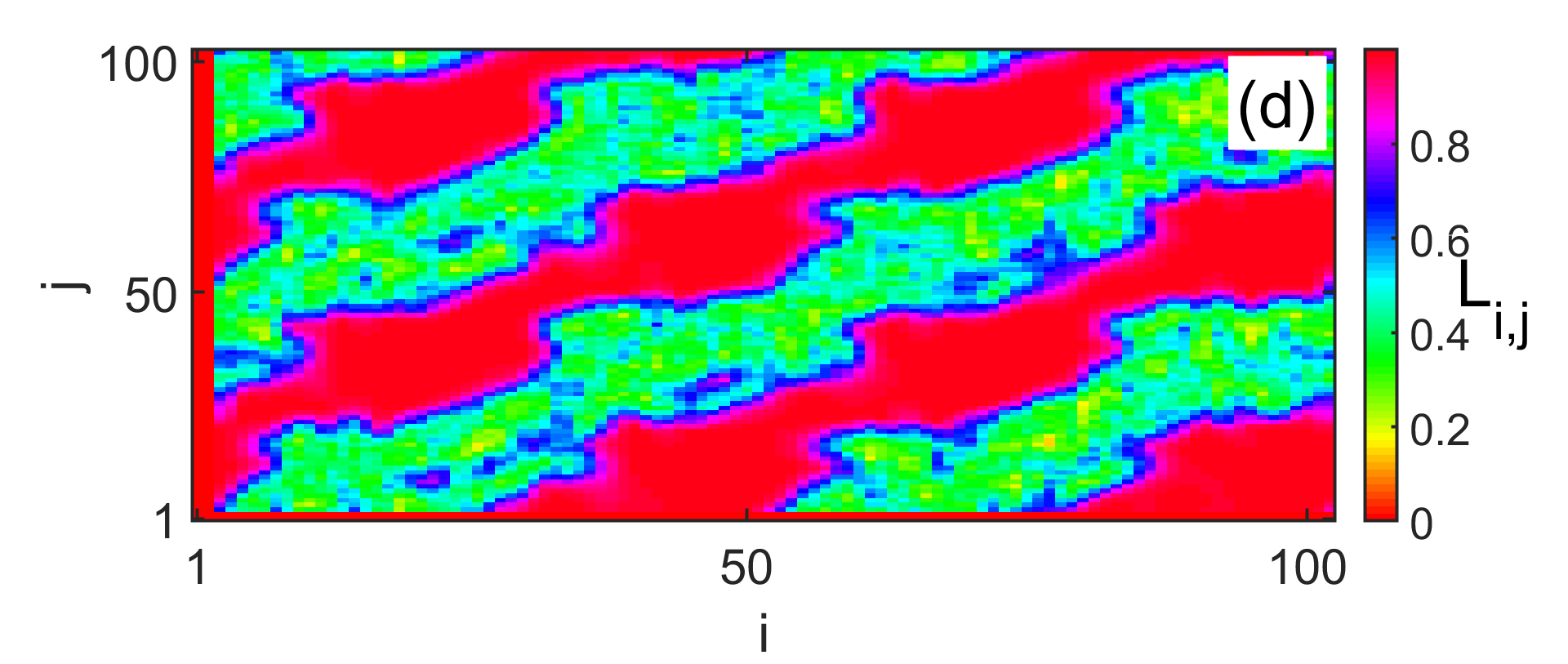}
	\caption{\label{fig.sgr7}The variation of the 2D local order parameter at a numerical time $t$ = 9700:  corresponding coupling strengths (a) $k_1=0$, $k_2=9$, (b) $k_1=1$, $k_2=1$, (c) $k_1=3$, $k_2=2$ and (d) $k_1=5$, $k_2=2.75$.}
\end{figure}

The variation of the coupling coefficients in order to have more information on their influences, allows us to show that, for the values of the electrical and chemical coupling coefficients respectively equal to 3 and 1 ($k_1 = 3$ and $k_2 = 1$), the traveling chimera  changes direction of propagation in time. This ability to move in one direction over a period of time and in the opposite direction in another during the evolution of the whole network, gives it a character of instability. Hence the notion of the alternating traveling chimera state in 2D network is observed. To illustrate this, we present in Fig. 6 a few images taken at different times chosen. It turns out that for the numerical time interval containing respectively the values $t=$ 3020, 3185, 3350, the patterns are oriented from left to right; Then comes a transition phase (time interval containing the values $t=$ 4670, 5660 and 6155) which allows the mutation of the sense of orientation of the patterns from right to left (interval containing the values 6320 and 6485). So in the opposite direction of propagation. And so on, the phenomenon of alternation of the sense of orientation of traveling chimera repeats itself intermittently as shown below. It is important to specify that the transitional phases during which we observe a progressive destruction of the traveling structure for the adoption of the opposite direction of orientation are not of long duration.

The presence of the traveling chimera and the clusters is clearly visible in the different figures. However, to ensure the existence of coherent states, we use the 2D local order parameter. This parameter allows us to specify whether a neuron belongs to a coherence zone or an incoherence zone depending on whether its value is equal (or very close) to 1 or even whether it is equal (or very close) to 0. Note that this parameter is calculated for each element in the 2D plane. For the computation of the 2D local order parameter $L_{i,k}$ by element $(i,k)$, requires to define a small grid of size $L \times L$ (with $L=\eta+1)$ of the neighboring elements of $(i,k)$ and centered in $(i,k)$. It is defined as follows,

\begin{equation}\label{eq.Lij.mt}L_{i,k}=\left|\dfrac{1}{(2\eta+1)^2}\sum_{\left|i-\alpha\right|\leq{\eta}}\sum_{\left|k-\beta\right|\leq{\eta}}\mathrm{e}^{\sqrt{-1}\Phi_{\alpha,\beta}}\right|,\end{equation}
where $\eta$ represents the number of neighbors to the left and right, to the bottom and top of neuron at the position ($i,k$). Here, $\Phi_{i,k}$ the geometric phase determined by,
\begin{equation}\label{eq.Phi.mt}\Phi_{i,k}=\arctan{(\dfrac{y_{i,k}}{x_{i,k}})}\end{equation}  and which is a good approximation as long as \textcolor{blue}{$r$} is small ( $\ll 1$ ) \cite{majhi2016chimera}.

The application of this measurement to the elements of our 2D network makes it possible to confirm that the suspected areas of coherence are true. We therefore notice in Fig. \ref{fig.sgr6} represented for a numerical time t = 9700 that the areas in red correspond well to values very close to 1 (red) as shown by the color bars. This testifies to the coherence of neurons in these areas. The multi-colored areas correspond to several values of the local order parameter also signify the incoherent state. We specify that the Figs. \ref{fig.sgr7}a, \ref{fig.sgr7}b, \ref{fig.sgr7}c and \ref{fig.sgr7}d, correspond respectively to the Fig. \ref{fig.sgr2}a, It allows us to confirm that these are indeed the phenomena of traveling chimera and multi-cluster.

\section{Network energy analysis}

Energy analysis is regularly used to understand the internal dynamics of elements under study. In this part, we use the study of the Hamiltonian to qualify the relative states of the elements in the 2D lattice in order to determine the direction of propagation of the traveling chimera.

Sarasola et al. \cite{sarasola2004} have proposed a mathematical formalism for the search of energy-like functions for well-known chaotic systems. This method is based on the decomposition of the speed vector $\dot{x}=f(x)$ of a dynamic system into two components, one of which is conservative ($f_c$), containing  the full rotation and the other dissipative ($f_d$), containing the divergence.
They show that, with the conservative field, the Hamiltonian function $H(x)$ can be calculated through the equation $\nabla H^T f_c (x)=0 $ which defines for each dynamical system a partial differential equation. Then the energy dissipated passively or actively by the dissipative components of the speed is given by the equation $\dot{H}=\nabla H^T f_d(x)$.

By applying this method to the Hindmarsh-Rose system represented by Eq.1, its energy satisfies the following equation,
\begin{eqnarray}(y -z)\frac{\partial H}{\partial x}-dx^2\frac{\partial H}{\partial y} + rsx\frac{\partial H}{\partial z}=0 \end{eqnarray} 
with a cubic polynomial solution \cite{sarasola2006},
\begin{equation}\label{eq.H.mt}H=[\frac{2}{3}dx^3 + rsx^2+(y-z)^2].\end{equation} 
    With this energy function, the derivative energies $\dot{H}=\nabla H^T f_d(x)$ become
 \begin{eqnarray}\label{eq.dh.mt}
 \dot{H}/2=(bd-rsx_0)x^4 +dIx^2-y^2-adx^5 \nonumber \\
 + (c+rsx_0)y-(c+rsx_0)z \nonumber \\
 +(1+r)yz+rsbx^3+rsIx-rz^2. 
 \end{eqnarray}

\begin{figure}[!h]
	\includegraphics[width=8.5cm]{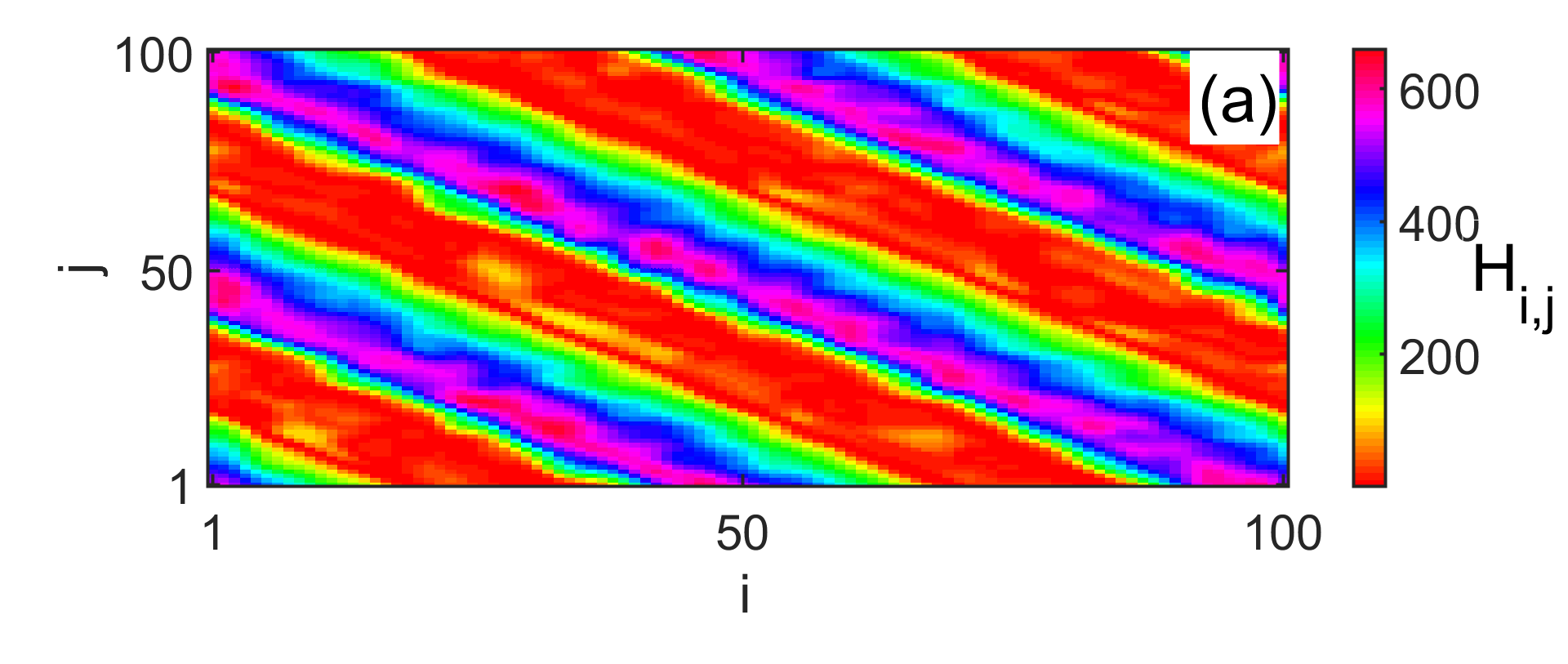}
	\includegraphics[width=8.5cm]{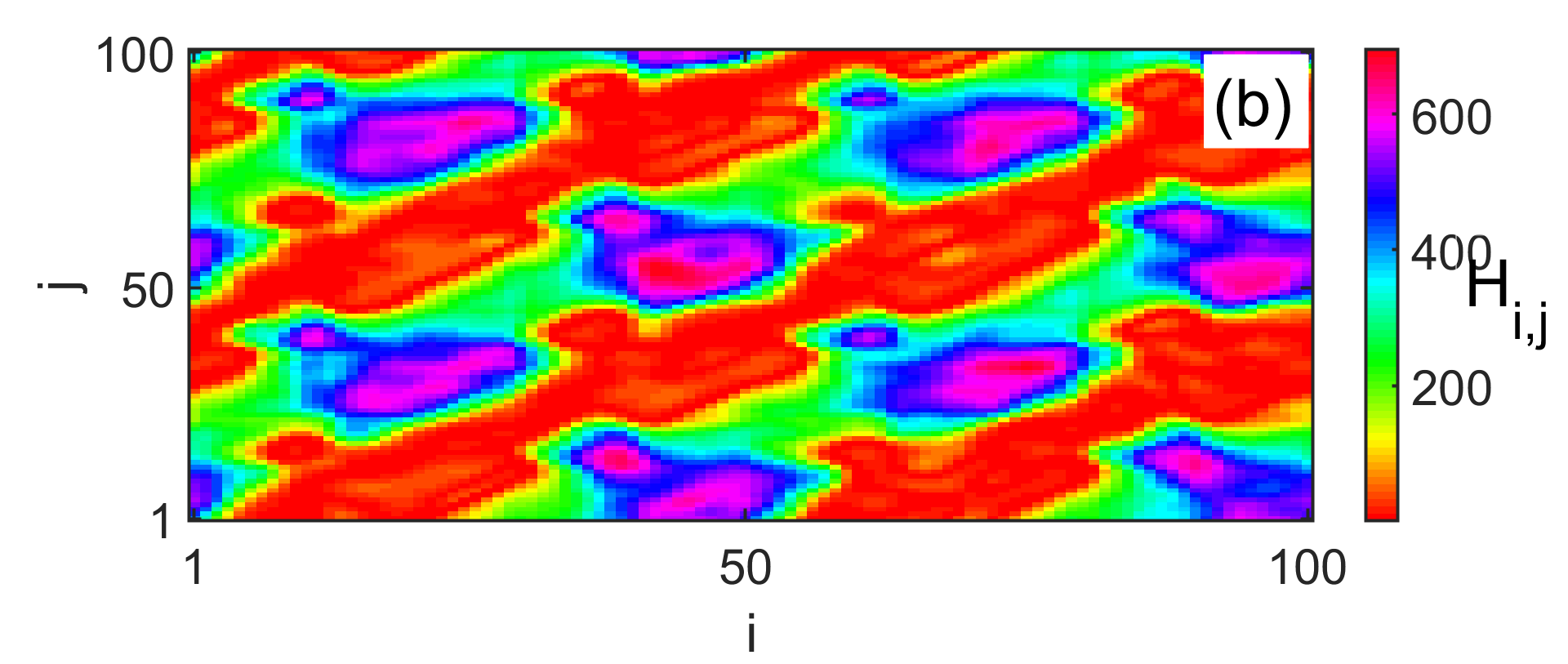}
	\includegraphics[width=8.5cm]{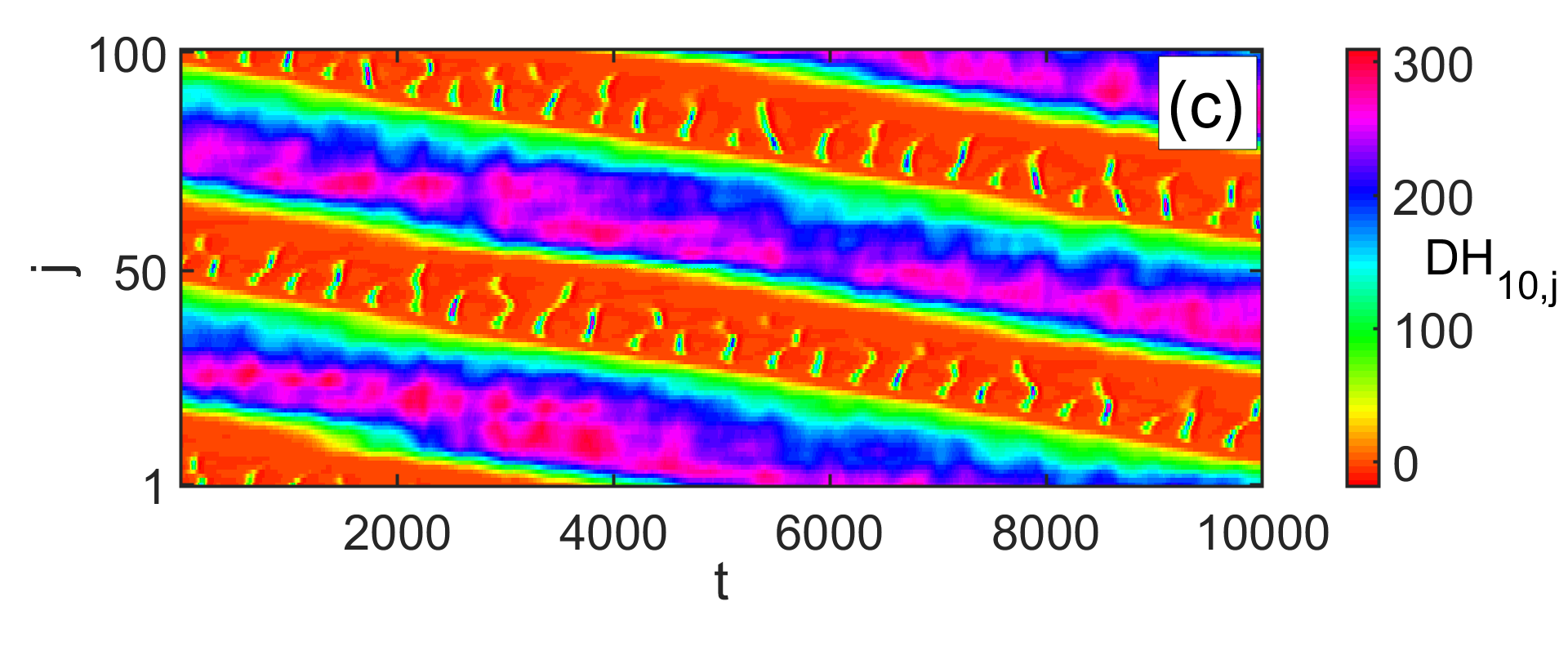}
	\includegraphics[width=8.5cm]{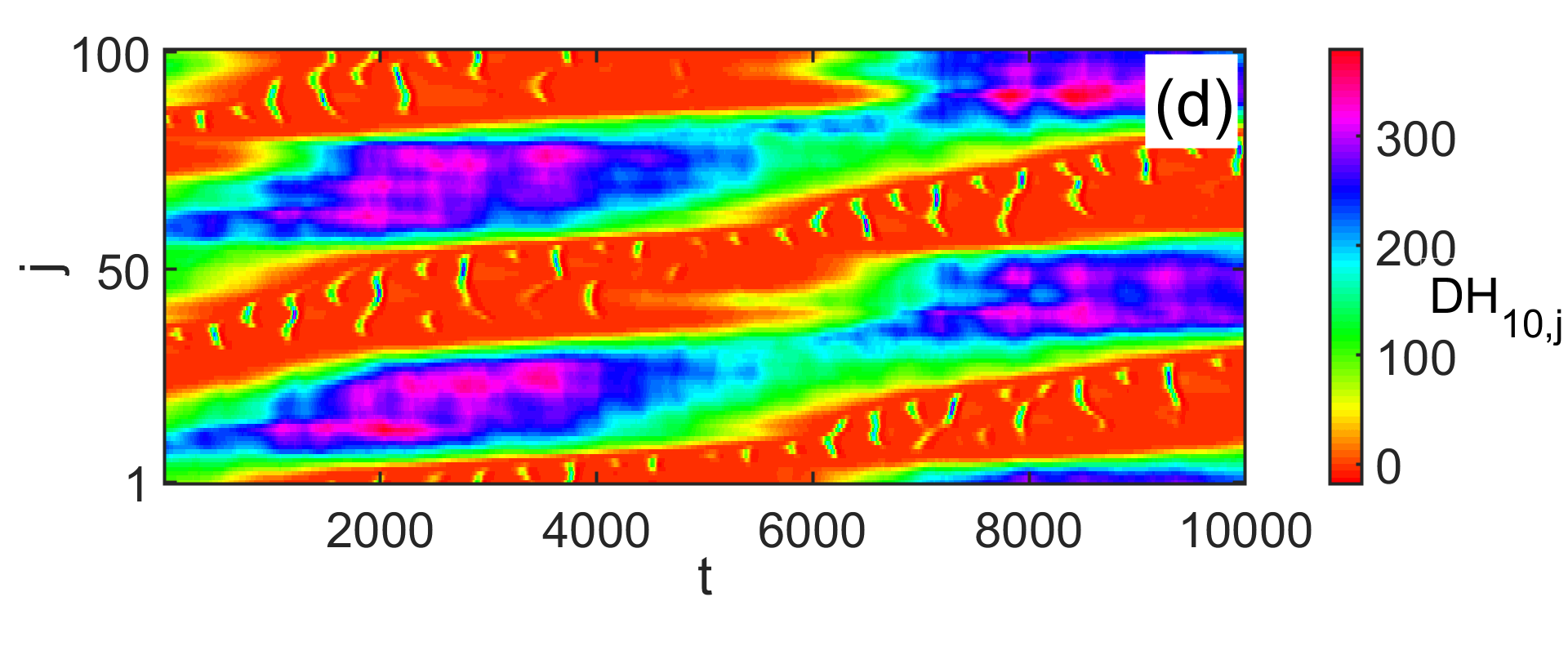}
	\caption{\label{fig.sgr8} Variation of energies in the 2D plane of the 2D neuronal network at $t$=9700,and the temporal evolution of the derivative energies $DH_{i, j, k}$ of the cross-section defined by ($i=10, j=1, ..., M$). (a, c) $k_1=5$, $k_2=2.5$, and (b, d) $k_1=5$, $k_2=2.75$.}
\end{figure}

We choose to evaluate these functions for all the elements of the network corresponding to the case where the values of the coupling coefficients are respectively $k_1 = 5,$ $k_2 = 2.5$ then $k_1 = 5$, $k_2 = 2.75$. The representation of the energy at a given instant (numerical time $t$ = 9700) shows that the elements belonging to the incoherent zones have higher energies than those belonging to the coherent zones (see Figs. \ref{fig.sgr8}(a) and \ref{fig.sgr8}(b)). Regarding the direction of propagation of the traveling chimera, we choose to evaluate the energy derivative of each element of a section t parallel to the x-axis (indices $i$) represented by $DH_{i, j, k}$ with fix values of $i$ and $k$. The elements of the section being numbered from the smallest index to a greater one ({\it i.e.,} from 1 to 100), it is easier to determine the direction of propagation of the chimera traveling by analyzing the direction of increase or decrease of the energy in time. This can be seen from the progression of the instantaneous derivative of the energy of each of the chain elements. Thus, for the cases of the first pair of coupling coefficients ($k_1 = 5, k_2 = 2.5$), it appears that the bands of high energy derivatives evolve in the opposite direction to that of the indices, {\it i.e.,} for a band of derivatives the indices decrease with time as shown in Fig. \ref{fig.sgr8}(c), unlike the case of the second pair of coupling coefficients ($k_1 = 5, k_2 = 2.75$) where the high energy derivative bands are rather oriented in the increasing direction of the indices of the line (Fig. \ref{fig.sgr8}(c)). A meticulous comparison to the displacement of the patterns over the entire 2D network allows us to specify that for the first pair of coefficients ($k_1 = 5, k_2 = 2.5$), this displacement goes from the right upper corner to bottom left in 2D grid (see Fig. \ref{fig.sgr5}(a)). while for the second case, it is done from the top left to the bottom the right corner of the bottom of the 2D grid (see Fig. \ref{fig.sgr5}(b)).

\section{Conclusion}

We have studied a 2D neuronal network composed of HR neurons. Our motivation is to extend the study conducted by Mishra et al. \cite{mishra2017traveling} to a more complex two-dimensional network and to check whether the same collective states could reproduce there. Starting from the same basic configuration including local electrical coupling and non-local chemical coupling between network's elements, we highlight the influence of coupling coefficients. First of all, we only consider chemical coupling. This leads to the organization of the 2D network into a grouping in the plane of the grid into bands of coherent elements interspersed by other bands of incoherent elements (Figs. 2(a) and 2(b)). Observing the evolution of these bands over time shows us that they are not static and move on the grid plane in both dimensions regularly and periodically (Fig. 2(c)). Hence the notion of \textit{traveling chimera state in 2D network} took place.

The next step was to introduce the electrical coupling coefficient. The mutual influence of the electrical and chemical coupling coefficients has presented us with a spectrum of behavior. It emerges that for certain values of coefficients of the two types of couplings, one expects the formation of clusters in displacement on the plane of the grid (Fig. 3) to which we have given the name of traveling multi-clusters; for others, a distortion of the patterns appears (Fig. 4); then, other parameters led to the change in the direction of propagation of these traveling patterns (Fig. 5) and later to an alternation in time of the traveling patterns, hence the name of \textit{alternating traveling chimera in 2D network} (Fig. 6).

To confirm the 2D coherent states, we have defined the two-dimensional local order parameter called 2D-local order parameter which allows to specify the membership for a given element to a coherence zone defined by a square centered in the latter. Then the use of the energy function proposed by Sarasola et al. \cite{sarasola2004} allowed us to determine the direction of propagation of the patterns in the grid. The implementation of this topology in a 3D network would be important for future studies because it would make it possible to generalize the dynamics of one dimensional and two dimensional networks to three dimensional networks in order to approach the complex connectivity brain network.

\section*{Acknowledgements}

HAC thanks ICTP-SAIFR and FAPESP grant
2016/01343-7 for partial support.


\bibliographystyle{prsty}

\end{document}